\documentclass[ALICE,manyauthors]{cernphprep}

\usepackage[comma,square,numbers,sort&compress]{natbib}
\usepackage{hyperref}
\usepackage{lineno}
\usepackage{color}
\usepackage{floatrow}
\usepackage[T1]{fontenc}

\begin{document}%


\newcommand{\ITS}          {\mathrm{ITS}}
\newcommand{\TOF}          {\mathrm{TOF}}
\newcommand{\ZDC}          {\mathrm{ZDC}}
\newcommand{\ZDCs}         {\mathrm{ZDCs}}
\newcommand{\ZNA}          {\mathrm{ZNA}}
\newcommand{\ZNC}          {\mathrm{ZNC}}
\newcommand{\SPD}          {\mathrm{SPD}}
\newcommand{\SDD}          {\mathrm{SDD}}
\newcommand{\SSD}          {\mathrm{SSD}}
\newcommand{\TPC}          {\mathrm{TPC}}
\newcommand{\VZERO}        {\mathrm{V0}}
\newcommand{\VZEROA}       {\mathrm{V0-A}}
\newcommand{\VZEROC}       {\mathrm{V0-C}}
\newcommand{\pip}          {$\pi^{+}$}
\newcommand{\pim}          {$\pi^{-}$}
\newcommand{\kap}          {K$^{+}$}
\newcommand{\kam}          {K$^{-}$}
\newcommand{\he}{$^{3}{\mathrm{He}}$}
\newcommand{\ahe}{$^{3}\overline{\rm He}$}
\newcommand{\ad}{$\overline{\rm d}$}
\newcommand{\pbar}         {$\mathrm{\overline{p}}$}
\newcommand{\kzero}        {\ensuremath{\mathrm{ K^{0}_{S}}}}
\newcommand{\vzero}        {\ensuremath{\mathrm{V}^0}}
\newcommand{\lmb}          {\ensuremath{\Lambda}}
\newcommand{\almb}         {\ensuremath{\bar{\Lambda}}}
\newcommand{\allpart}      {$\pi^{\pm}$, K$^{\pm}$, \kzero, p(\pbar) and \lmb(\almb)}
\newcommand{\allpikp}      {$\pi^{\pm}$, K$^{\pm}$ and p(\pbar)}
\newcommand{\pikp}         {$\pi$, K, and p}
\newcommand{\allpi}        {$\pi^{\pm}$}
\newcommand{\allk}         {K$^{\pm}$}
\newcommand{\allp}         {p(\pbar)}
\newcommand{\alllmb}       {\lmb(\almb)}
\newcommand{\degree}       {$^{\mathrm{o}}$}
\newcommand{\dg}           {\mbox{$^\circ$}}
\newcommand{\dedx}         {\ensuremath{\mathrm{d}E/\mathrm{d}x}}
\newcommand{\dndy}         {d$N$/d$y$}
\newcommand{\pp}           {pp}
\newcommand{\ppbar}        {\mbox{$\mathrm {p\overline{p}}$}}
\newcommand{\PbPb}         {\mbox{Pb--Pb}}
\newcommand{\pPb}          {\mbox{p--Pb}}
\newcommand{\AuAu}         {\mbox{Au--Au}}
\newcommand{\pseudorap}    {\mbox{$\left | \eta \right | $}}
\newcommand{\dNdeta}       {\ensuremath{\mathrm{d}N_\mathrm{ch}/\mathrm{d}\eta}}
\newcommand{\dNdy}         {\ensuremath{\mathrm{d}N_\mathrm{ch}/\mathrm{d}y}}
\newcommand{\dNdyst}       {\ensuremath{\sqrt{\frac{dN_\pi/dy}{s_T}}}}
\newcommand{\dNdetatr}     {\mathrm{d}N_\mathrm{tracklets}/\mathrm{d}\eta}
\newcommand{\dNdetar}[1]   {\mathrm{d}N_\mathrm{ch}/\mathrm{d}\eta\left.\right|_{|\eta|<#1}}
\newcommand{\dNdetamean}   {\ensuremath{\langle\dNdeta  \rangle}}
\newcommand{\lum}          {\, \mbox{${\mathrm{ cm}}^{-2} {\mathrm {s}}^{-1}$}}
\newcommand{\barn}         {\, \mbox{${\mathrm{barn}}$}}
\newcommand{\m}            {\, \mbox{${\mathrm{m}}$}}
\newcommand{\ncls}         {\ensuremath{N_{cls}}}
\newcommand{\nsigma}       {\ensuremath{n\sigma}}
\newcommand{\dcaxy}        {\ensuremath{\mathrm{DCA_{\rm xy}}}}
\newcommand{\dcaz}         {\ensuremath{\mathrm{DCA_{\rm z}}}}
\newcommand{\EcrossB}      {E$\times$B}
\newcommand{\bb}           {Bethe-Bloch}
\newcommand{\s}            {\ensuremath{\sqrt{s}}}
\newcommand{\PT}           {\ensuremath{p_{\mathrm{T}}}}
\newcommand{\MT}           {\ensuremath{m_{\mathrm{T}}}}
\newcommand{\hlab}         {\ensuremath{\eta_{\mathrm{lab}}}}
\newcommand{\ynn}         {\ensuremath{y_{\mathrm{NN}}}}
\newcommand{\ycms}         {\ensuremath{y_{\mathrm{CMS}}}}
\newcommand{\ylab}         {\ensuremath{y_{\mathrm{lab}}}}
\newcommand{\ppi}          {\ensuremath{{\mathrm{p}}/\pi}}
\newcommand{\kpi}          {\ensuremath{{\mathrm{K}}/\pi}}
\newcommand{\lpi}          {\ensuremath{{\mathrm{ \Lambda}}/\pi}}
\newcommand{\lks}          {\ensuremath{{\mathrm{ \Lambda}}/\mathrm{{K}^{0}_{S}}}}
\newcommand{\mt}           {\ensuremath{m_{\mathrm{T}}}}
\newcommand{\snn}          {\ensuremath{\sqrt{s_{\mathrm{NN}}}}}
\newcommand{\snnbf}        {\ensuremath{\mathbf{{\sqrt{s_{\mathbf{ NN}}}}}}}
\newcommand{\sonly}        {\ensuremath{\sqrt{s}}}
\newcommand{\Npart}        {\ensuremath{N_\mathrm{part}}}
\newcommand{\avNpart}      {\ensuremath{\langle N_\mathrm{part} \rangle}}
\newcommand{\avNpartdata}  {\ensuremath{\langle N_\mathrm{part}^{\mathrm{data}} \rangle}}
\newcommand{\Ncoll}        {\ensuremath{N_\mathrm{coll}}}
\newcommand{\avNcoll}      {\ensuremath{\langle N_\mathrm{coll} \rangle}}
\newcommand{\Dnpart}       {\ensuremath{D\left(\Npart\right)}}
\newcommand{\DnpartExp}    {\ensuremath{D_{\mathrm{exp}}\left(\Npart\right)}}
\newcommand{\dNdetapt}     {\ensuremath{\dNdeta\,/\left(0.5\Npart\right)}}
\newcommand{\dNdetaptr}[1] {\ensuremath{\dNdetar{#1}\,/\left(0.5\Npart\right)}}
\newcommand{\dNdetape}     {\left(\ensuremath{\dNdeta\right)/\left(\avNpart/2\right)}}
\newcommand{\dNdetaper}[1] {\ensuremath{\dNdetar{#1}\,/\left(\avNpart/2\right)}}
\newcommand{\dndydpt}      {\ensuremath{{\mathrm{d}}^2N/({\mathrm{d}}y {\mathrm{d}}p_{\mathrm{t}})}}
\newcommand{\abs}[1]       {\ensuremath{\left|#1\right|}}
\newcommand{\signn}        {\ensuremath{\sigma^{\mathrm{inel.}}_{\mathrm{NN}}}}
\newcommand{\vz}           {\ensuremath{V_{z}}}
\newcommand{\Tfo}          {\ensuremath{{T}_{\mathrm{kin}}}}
\newcommand{\Tch}          {\ensuremath{{T}_{\mathrm{ch}}}}
\newcommand{\bT}           {\ensuremath{\beta_{\mathrm{T}}}}
\newcommand{\avbT}         {\ensuremath{\langle \beta_{\mathrm{T}}\rangle}}
\newcommand{\avpT}         {\ensuremath{\langle \PT \rangle}}
\newcommand{\muB}          {\ensuremath{\mu_{B}}}
\newcommand{\stat}         {({\it stat.})}
\newcommand{\syst}         {({\it sys.})}
\newcommand{\Fig}[1]       {Fig.~\ref{#1}}
\newcommand{\Ref}[1]       {Ref.~\cite{#1}}
\newcommand{\green}[1]     {\textcolor{green}{#1}}
\newcommand{\blue}[1]      {\textcolor{blue}{#1}}
\newcommand{\red}[1]       {\textcolor{red}{#1}}
\newcommand{\white}[1]     {\textcolor{white}{#1}}
\newcommand{\gevc}         {\ensuremath{{\mathrm{GeV}}/c}}
\newcommand{\gevcsq}       {\ensuremath{{\mathrm{GeV}^4}/c^2}}
\newcommand{\mevc}         {\ensuremath{{\mathrm{MeV}}/c}}
\newcommand{\avg}[1]       {\ensuremath{\left\langle#1\right\rangle}}
\newcommand{\tkin}         {\ensuremath{T\mathrm{_{kin}}}}
\newcommand{\rppb}         {\ensuremath{R\mathrm{_{pPb}}}}
\newcommand{\rdau}         {\ensuremath{R\mathrm{_{dAu}}}}
\newcommand{\mtof}         {\ensuremath{{m_{\mathrm{TOF}}^2}}}
\newcommand{\dbard}        {\ensuremath{\mathrm{\bar{d}/d}}}
\newcommand{\pbarp}        {\ensuremath{\mathrm{\bar{p}/p}}}
\newcommand{\nbarn}        {\ensuremath{\mathrm{\bar{n}/n}}}
\newcommand{\alphas}       {\ensuremath{\alpha_{\mathrm{S}}}}
\newcommand{\mub}          {\ensuremath{\mathrm{\mu}_{\mathrm{B}}}}
\newcommand{\Btwo}         {\ensuremath{B_2}}

\newcommand{\warn}[1]      {{\small\textbf{\textcolor{red}{(!\footnote{\textbf{(!)}~#1})}}}}
\newcommand{\warnin}[1]         {\textit{\textcolor{red}{(#1)}}}
\newcommand{\fake}[1]      {\textbf{\textcolor{red}{#1}}}
\newcommand{\final}[1]     {\textbf{\textcolor{blue}{#1}}}
\newcommand{\prelim}[1]    {\textbf{\textcolor{magenta}{#1}}}
\renewcommand{\mod}[1]       {\textbf{\textcolor{red}{#1}}}

\begin{titlepage}
%
\PHyear{2019}
\PHnumber{120}      
\PHdate{31 May}  

\title{Multiplicity dependence of light (anti-)nuclei production\\ in p--Pb collisions at $\mathbf{\snn}$~=~5.02~TeV}
\ShortTitle{Production of (anti-)nuclei in p--Pb collisions} 

\Collaboration{ALICE Collaboration\thanks{See Appendix~\ref{app:collab} for the list of collaboration members}}
\ShortAuthor{ALICE Collaboration} 

\begin{abstract}

The measurement of the deuteron and anti-deuteron production in the rapidity range $-1 < y < 0$ as a function of transverse momentum and event multiplicity in \pPb{} collisions at \snn{} = 5.02~TeV is presented. \mbox{(Anti-)}deuterons are identified via their specific energy loss \dedx{} and via their time-of-flight. Their production in p--Pb collisions is compared to pp and Pb--Pb collisions and is discussed within the context of thermal and coalescence models. The ratio of integrated yields of deuterons to protons (d/p) shows a significant increase as a function of the charged-particle multiplicity of the event starting from values similar to those observed in pp collisions at low multiplicities and approaching those observed in Pb--Pb collisions at high multiplicities. The mean transverse particle momenta are extracted from the deuteron spectra and the values are similar to those obtained for p and $\Lambda$ particles. 
Thus, deuteron spectra do not follow mass ordering.
This behaviour is in contrast to the trend observed for non-composite particles in p--Pb collisions. In addition, the production of the rare \he\ and \ahe{} nuclei has been studied. The spectrum corresponding to all non-single diffractive p-Pb collisions is obtained in the rapidity window $-1 < y < 0$ and the \PT-integrated yield \dndy\ is extracted. It is found that the yields of protons, deuterons, and \he{}, normalised by the spin degeneracy factor, follow an exponential decrease with mass number. 

\end{abstract}
\end{titlepage}
\setcounter{page}{2}

\section{Introduction}
The energy densities reached in the collisions of ultra-relativistic particles lead to a significant production of complex \mbox{(anti-)}\mbox{(hyper-)}nuclei. The high yield of anti-quarks produced in these reactions has led to the first observation of the anti-alpha particle~\cite{Agakishiev:2011ib} as well as of the anti-hyper-triton~\cite{Abelev:2010} by the STAR collaboration, and to detailed measurements by the ALICE collaboration \cite{nuclei,Adam:2015yta,Acharya:2017dmc,Acharya:2019rgc} at energies reached at the CERN LHC. However, the production mechanism is not fully understood. In a more general context, these measurements also provide input for the background determination in searches for anti-nuclei in space. Such an observation of anti-deuterons or \ahe\ of cosmic origin could carry information on the existence of large amounts of anti-matter in our universe or provide a signature of the annihilation of dark matter particles~\cite{Blum:2017qnn,Poulin:2018wzu,Tomassetti:2017qjk,Korsmeier:2017xzj,Cui:2010ud}.

Recent data in pp and in heavy-ion collisions provide evidence for an interesting observation regarding the production mechanism of \mbox{(anti-)}nuclei~\cite{nuclei,Acharya:2017fvb,Acharya:2017bso,Acharya:2017dmc,Acharya:2019rgc}: in \PbPb{} interactions, the d/p ratio does not vary with the collision centrality and the value agrees with expectations from thermal-statistical models which feature a common chemical freeze-out temperature of all hadrons around 156 MeV~\cite{nuclei,Andronic:2010qu,Cleymans:2011pe}.
In inelastic pp collisions, the corresponding ratio is a factor 2.2 lower than in Pb--Pb collisions \cite{nuclei, Acharya:2017fvb}. With respect to these measurements, the results of d and $^3$He produced in \pPb{} collisions at \snn{} = 5.02~TeV, being a system in between the two extremes of pp and Pb--Pb collisions, are of prominent interest and they are the subject of this letter. 
While deuterons have been measured differentially in multiplicity, the \ahe{} (\he){} spectrum was only obtained inclusively for all non-single diffractive events because of their low production rate.

In addition to the evolution of the integrated d/p ratio for various multiplicity classes, the question whether the transverse momentum distribution of deuterons is consistent with a collective radial expansion together with the non-composite hadrons is of particular interest. Such behaviour has been observed for light nuclei in \PbPb{} collisions~\cite{nuclei,Acharya:2017dmc}. The presence of collective effects in \pPb\ collisions at LHC energies has recently been supported by several experimental findings (see for instance~\cite{Abelev:2012ola,ABELEV:2013wsa,Abelev:2013haa,CMS:2012qk,Aad:2012gla,Aad:2013fja,Chatrchyan:2013nka} and recent reviews in~\cite{Loizides:2016tew,Nagle:2018nvi}). 
These include a clear mass ordering of the mean transverse momenta of light flavoured hadrons in p--Pb collisions as expected from hydrodynamical models~\cite{Abelev:2013haa}.

\section{Analysis}

The results presented here are based on a low pile-up p--Pb data sample collected with the ALICE detector during the LHC running campaign at \snn~=~5.02~TeV in 2013. A detailed description of the detector is available in~\cite{Alessandro:2006yt,Carminati:2004fp,Aamodt:2008zz,Aamodt:2010dx,Abelev:2014ffa}. The main detectors used in this analysis are the Inner Tracking System (ITS)~\cite{Aamodt:2010aa}, the Time Projection Chamber (TPC)~\cite{Alme:2010ke}, and the Time-Of-Flight detector (TOF)~\cite{Akindinov:2010zzb,Akindinov:2013tea}. The two innermost layers of the ITS consist of Silicon Pixel Detectors (SPD), followed by two layers of Silicon Drift Detectors (SDD), and two layers of Silicon Strip Detectors (SSD). As the main tracking device, the TPC provides full azimuthal acceptance for tracks in the pseudo-rapidity region $|\hlab| <$ 0.8. In addition, it provides particle identification via the measurement of the specific energy loss d$E$/d$x$.
The TOF array is located at about 3.7 m from the beam line and provides particle identification by measuring the particle speed with the time-of-flight technique.
In p-Pb collisions, the overall time resolution is about 85~ps for high multiplicity events.
In peripheral events, where multiplicities are similar to pp, it decreases to about 120~ps due to a worse start-time (collision-time) resolution~\cite{Adam:2016ilk}. All detectors are positioned in a solenoidal magnetic field of $B$ = 0.5~T.

The  event  sample  used  for  the  analysis  presented  in  this  letter 
was  collected exclusively in the beam configuration where the proton travels towards negative \hlab. The minimum-bias trigger signal and the definition of the multiplicity classes was provided by the V0 detector consisting of two arrays of 32 scintillator tiles each covering the full azimuth within $2.8 < \hlab < 5.1$ (V0A, Pb-beam direction) and $-3.7 < \hlab < -1.7$ (V0C, p-beam direction). The event selection was performed in a similar way to that described in Ref.~\cite{Abelev:2013haa}. A coincidence of signals in both V0A and V0C was required online in order to remove background from single diffractive and electromagnetic events. In the offline analysis, further background suppression was achieved by requiring that the arrival time of the signals in the two neutron Zero Degree Calorimeters (ZDC), which are located $\pm$112.5~m from the interaction point, is compatible with a nominal \pPb{} collision. 
The contamination from pile-up events was reduced to a negligible level ($<1\%$) by rejecting events in which more than one primary vertex was reconstructed either from SPD tracklets or from tracks reconstructed in the whole central barrel.
The position of the reconstructed primary vertex was required to be located within $\pm 10$~cm of the nominal interaction point in the longitudinal direction. In total, an event sample of about 100 million minimum-bias (MB) events after all selections was analysed. The corresponding integrated luminosity, $L_{\mathrm{int}} = N_\mathrm{MB}/\sigma_\mathrm{MB}$, where $\sigma_\mathrm{MB}$ is the MB trigger cross-section measured with van-der-Meer scans, amounts to 47.8~$\mu\mathrm{b}^{-1}$ with a relative uncertainty of 3.7\%~\cite{Abelev:2014epa}.

The final results are given normalised to the total number of non-single diffractive (NSD) events. Therefore, a correction of  3.6\%$\pm$3.1\% \cite{ALICE:2012xs} is applied to the minimum-bias results, which corresponds to the trigger and vertex reconstruction inefficiency for this selection. For the study of d and \ad{}, the sample is divided into five multiplicity classes, which are defined as percentiles of the V0A signal. This signal is proportional to the charged-particle multiplicity in the corresponding pseudo-rapidity region in the direction of the Pb-beam. 
Following the approach in~\cite{Adam:2016dau}, the multiplicity dependent results are normalized to the number of events $N_{\rm ev}$ corresponding to the visible (triggered) cross-section. The event sample is corrected for the vertex reconstruction efficiency. This correction is of the order of 4\% for the lowest V0A multiplicity class (60-100\%) and negligible ($<$1\%) for the other multiplicity classes. 
The chosen selection and the corresponding charged-particle multiplicity at mid-rapidity are summarized in Table~\ref{tab:mult-bins}.

\begin{table}[hb]
  \centering
  \begin{tabular}[hb]{rc}
       \hline
       \hline
    V0A Class &  $\left. \langle {\rm d}N_{\rm ch}/{\rm d}\eta_{\rm lab}  \rangle \right|_{\left|\eta_{\rm lab}\right| < 0.5}$  \\
    \hline 
    0--10\%   &  40.6  $\pm$ 0.9  \\
    10--20\%  &  30.5   $\pm$ 0.7 \\
    20--40\%  &  23.2   $\pm$ 0.5  \\
    40--60\%  &  16.1   $\pm$ 0.4  \\
    60--100\%  & 7.1   $\pm$ 0.2   \\
    \hline
    \hline
   \end{tabular}
   \caption{Multiplicity intervals and the corresponding charged-particle multiplicities at mid-rapidity. The uncertainties reported for the $\langle {\rm d}N_{\rm ch}/{\rm d}\eta_{\rm lab}  \rangle |_{\left|\eta_{\rm lab}\right| < 0.5}$ are the systematic ones, statistical uncertainties are negligible. Values are taken from \cite{Abelev:2013haa}.}
   \label{tab:mult-bins}
\end{table}

In this analysis, the production of primary deuterons and \he{}-nuclei and that of their respective anti-particles are measured in a rapidity window $-1 < y < 0$ in the centre-of-mass system. Since the energy per nucleon of the proton beam is higher than that of the Pb beam, the nucleon-nucleon system moves in the laboratory frame with a rapidity of -0.465.
 Potential differences of the spectral shape or normalisation due to the larger $y$-range with respect to the measurement of $\pi$, K,  and p \cite{Abelev:2013haa} are found to be negligible for the (anti-)deuteron and \he\ minimum-bias spectra  with respect to the overall statistical and systematic uncertainties.
In order to select primary tracks of suitable quality, various track selection criteria are applied. At least 70 clusters in the TPC and two hits in the ITS (out of which at least one in the SPD) are required. These selections guarantee a track momentum resolution of 2\% in the relevant \PT-range and a d$E$/d$x$ resolution of about~6\% for minimum ionising particles. The maximum allowed Distance-of-Closest-Approach (DCA) to the primary collision vertex is 
0.12~cm in the transverse (\dcaxy) and 1.0~cm in the longitudinal (\dcaz) plane. Furthermore, it is required that the  $\chi^2$ per TPC cluster is less than 4 and tracks of weak-decay products with kink topology are rejected \cite{Abelev:2014ffa}, as they cannot originate from the tracks of primary nuclei.

\begin{figure}[htbp]
  \begin{center}
      \includegraphics[trim=0cm 0cm 0cm 0cm, clip=true, width = 0.49\textwidth]{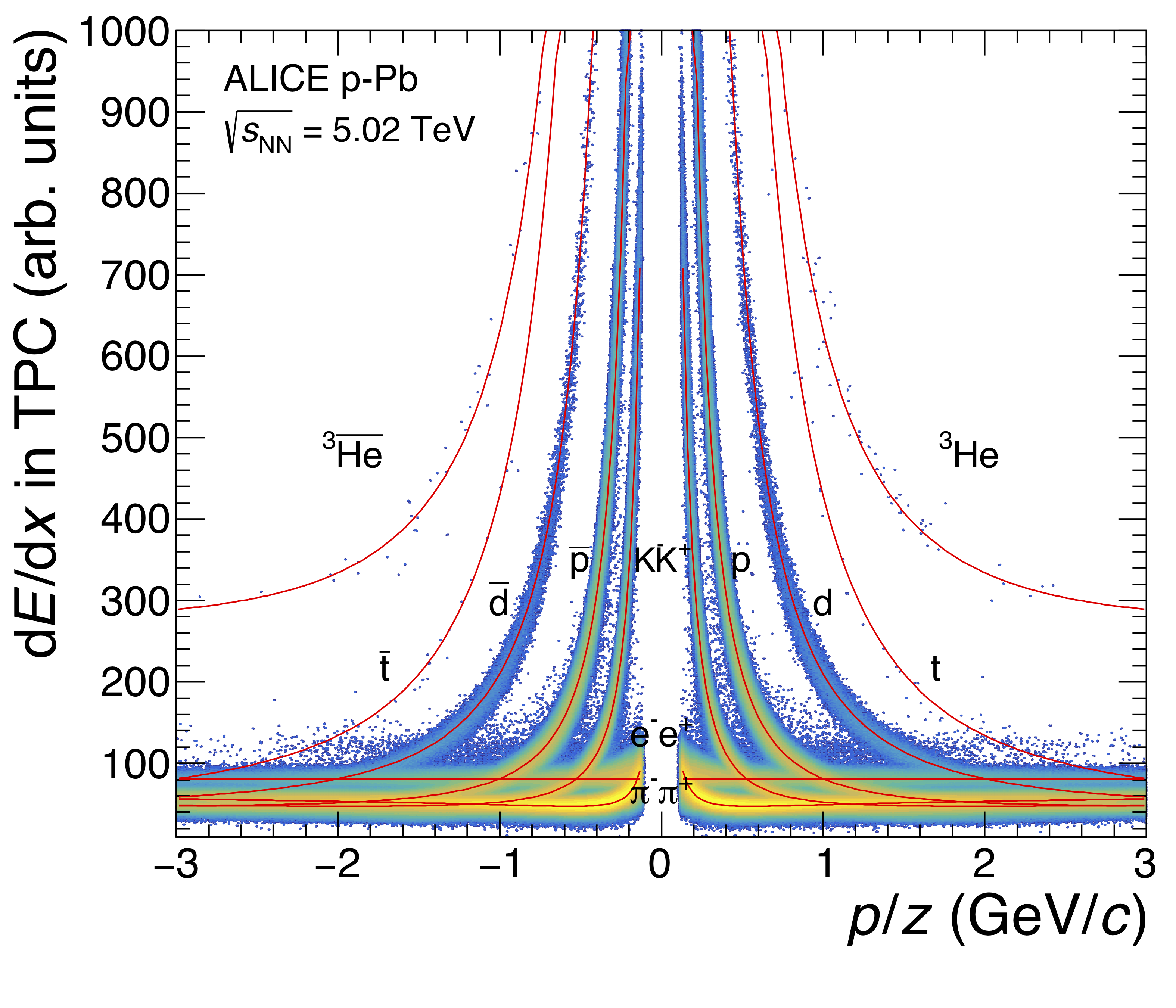} 
    \includegraphics[trim=0cm 0cm 0cm 0cm, clip=true, width = 0.49\textwidth]{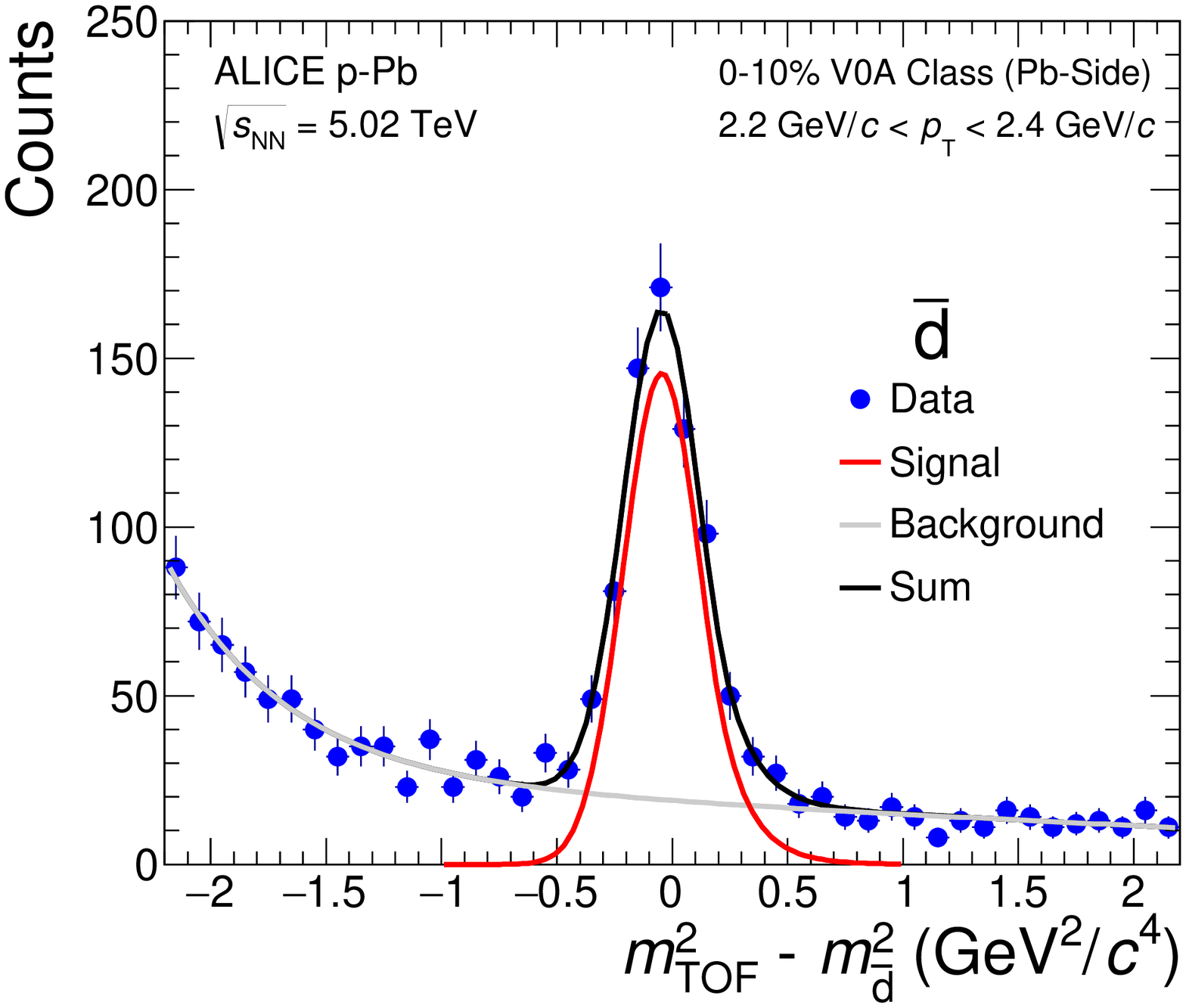} 
        \caption{Energy loss \dedx{} in the TPC and the corresponding expected energy loss from a parametrization of the Bethe-Bloch curve (left). Example of the fit to the squared TOF mass difference which shows separately the signal and the background from the exponential tail of protons and from mismatched tracks (right).}
    \label{fig:tpctofperf}
  \end{center}
\end{figure}

The particle identification performance of the TPC and TOF detectors in \pPb{} collisions is shown in Fig.~\ref{fig:tpctofperf}. For the mass determination with the TOF detector, the contribution of tracks with a wrongly assigned TOF cluster is largely reduced by a 3$\sigma$ pre-selection in the TPC d$E$/d$x$, where $\sigma$ corresponds to the TPC d$E$/d$x$ resolution. Nevertheless, due to the small abundance of deuterons the background is still significant and it is removed using a fit to the squared mass distribution. An example of a fit for anti-deuterons with transverse momenta $2.2~\textrm{GeV}/c < \PT{} < 2.4~\textrm{GeV}/c$ is shown in the right panel of Fig.~\ref{fig:tpctofperf}. The squared rest mass of the deuteron has been subtracted to simplify the fitting function. The signal has a Gaussian shape with an exponential tail on the right side. This tail is necessary to describe the time-signal shape of the TOF detector~\cite{Akindinov:2013tea}. For the background, the sum of two exponential functions is used. One of the exponential functions accounts for the mismatched tracks and the other accounts for the tail of the proton peak. For \mbox{(anti-)}\he{} nuclei, the \dedx{} is sufficient for a clean identification using only this technique over the entire momentum range $1.5~\gevc<$~\PT{} $< 5~\gevc$ as the atomic number $Z=2$ for $^3$He leads to a clear separation from other particles.

\begin{figure}[ht]
  \begin{center}
    \includegraphics[width = 0.49\textwidth]{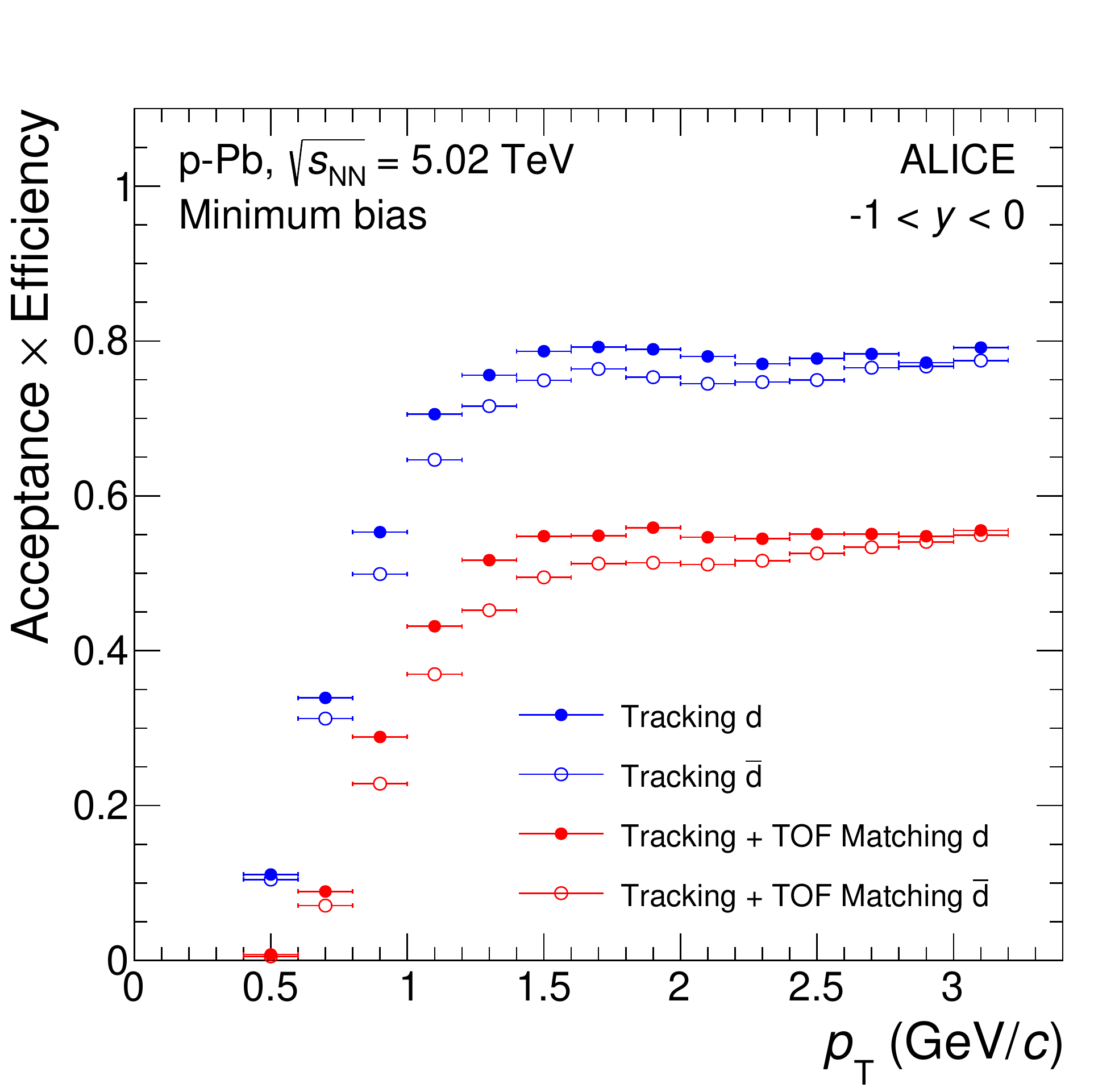} 
    \includegraphics[width = 0.49\textwidth]{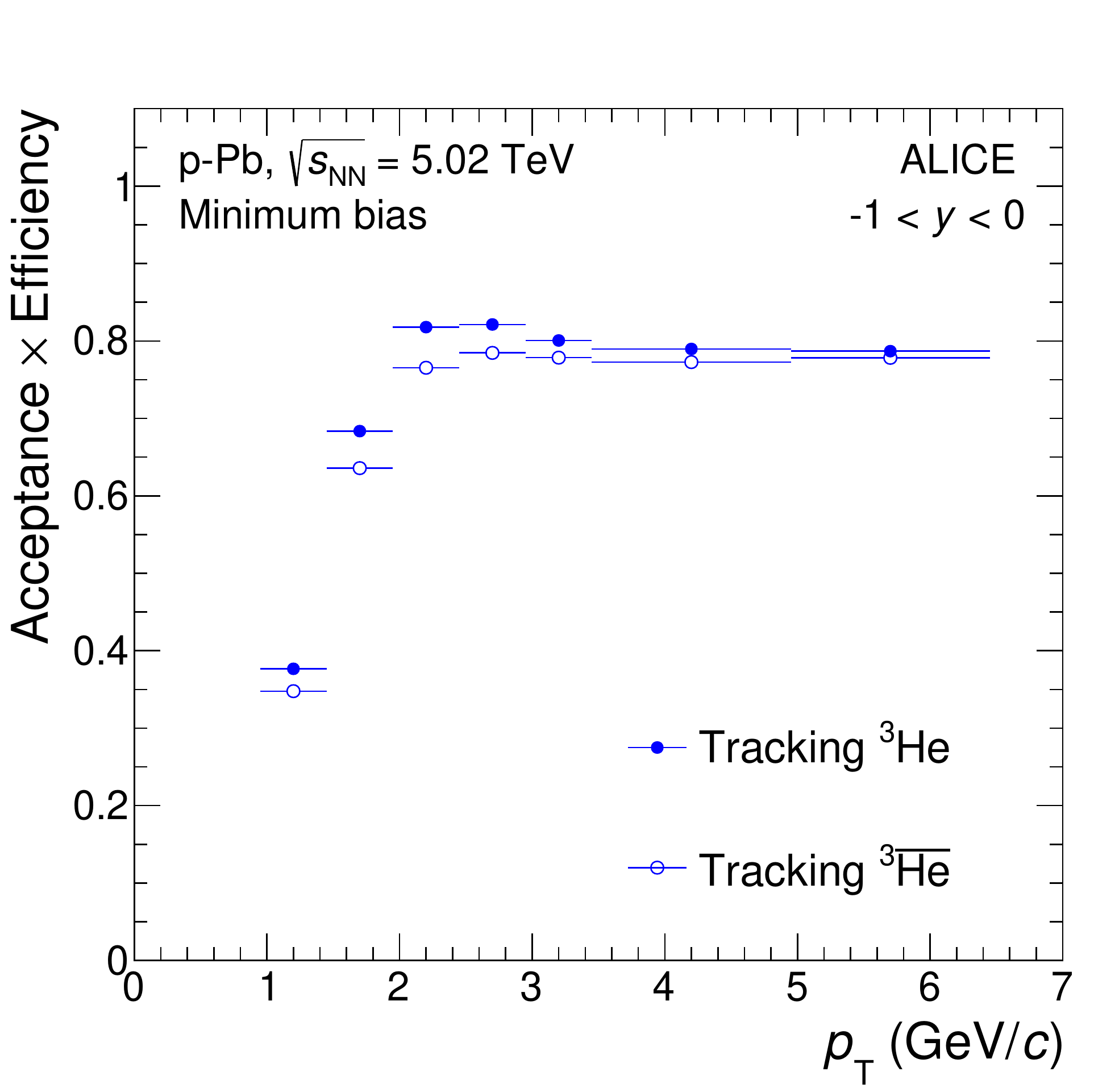}
    \caption{Tracking acceptance $\times$ efficiency correction for (anti-)deuterons (left) and for $^3$He and $^{3}\overline{\rm He}$ (right)  in the minimum-bias class. 
    The efficiencies for anti-nuclei are lower due to the larger cross-section for hadronic interactions.}
    \label{fig:eff}
  \end{center}
\end{figure}

The tracking acceptance $\times$ efficiency determination is based on a Monte-Carlo simulation using the DPMJET event generator~\cite{Roesler:2000he} and a full detector description in GEANT3~\cite{Geant:1994zzo}. As discussed in~\cite{nuclei}, the hadronic interaction of \mbox{(anti-)}nuclei with detector material is not fully described in GEANT3, therefore two additional correction factors are applied. Firstly, in order to account for the material between the collision vertex and the TPC, the track reconstruction efficiencies extracted from GEANT3 are scaled to match those from GEANT4~\cite{Agostinelli:2002hh,Uzhinsky:2011zz}. Secondly, for tracks which cross in addition the material between the TPC and the TOF detectors, a data-driven correction factor has been evaluated by comparing the matching efficiency of tracks to TOF hits in data and Monte Carlo simulation. Since the TRD was not fully installed in 2013, this study was repeated for regions in azimuth with and without installed TRD modules. The matching efficiencies for tracks crossing the TRD material were then scaled such that the corrected yield agrees with the one obtained for tracks that are not crossing any TRD material. This procedure results in a further reduction of the acceptance $\times$ efficiency of 6\% for deuterons and 11\% for anti-deuterons. 
The acceptance and efficiency corrections are found to be independent of the event multiplicity and are shown in Fig.~\ref{fig:eff} for primary deuterons and anti-deuterons, with and without requiring a TOF match, as well as for \he{} and \ahe{}.

%
%

The raw yields of deuterons and \he{} also include secondary particles which stem 
from the interactions of primary particles with the detector material.
To subtract this contribution, a data-driven approach as in~\cite{Abelev:2013haa,nuclei} is used. The distribution of the \dcaxy{} is fitted with two distributions (called "templates" in the following) obtained from Monte-Carlo simulations describing primary and secondary deuterons, respectively. The fit is performed in the range $|\dcaxy| < 0.5$~cm which allows the contribution from material to be constrained by the plateau of the distribution at larger distances ($|\dcaxy| > 0.15$~cm). The contamination of secondaries amounts to about 45\% to 55\% in the lowest \PT-interval and decreases exponentially towards higher \PT{} until it becomes negligible ($<$ 1\%) above 2~\gevc{}. The limited number  
of \he{} candidate tracks does not allow a background subtraction based on templates, instead a bin counting procedure in the aforementioned \dcaxy{} signal and background regions is used.

The systematic uncertainties of the measurement are summarised for deuterons and \he{} as well as for their antiparticles in Table~\ref{tab:systematics}. For deuterons, the uncertainty related to the secondary correction is estimated by repeating the template fit procedure under a variation of the \dcaz{} cut. The corresponding uncertainty for \he{} nuclei is determined by varying the ranges in \dcaxy{} for the signal and background regions in the bin counting procedure.
For d and \he{} the systematic uncertainty on the cross-section for hadronic interaction is determined by a systematic comparison of different propagation codes (GEANT3 and GEANT4).
The material between TPC and TOF needs to be considered only for the (anti-)deuteron spectrum and increases the uncertainty by additional 3\% and 5\% for deuterons and anti-deuterons, respectively. This corresponds to the half of the observed discrepancy in the TPC-TOF matching efficiencies evaluated in data and Monte Carlo.
For both deuterons and anti-deuterons, the particle identification procedure introduces only a small uncertainty which slightly increases at high \PT{} and is estimated based on the variation of the $n\sigma$-cuts in the TPC d$E$/d$x$ as well as on a variation of the signal extraction in the TOF with different fit functions. The PID related uncertainties for \he{} and \ahe{} remain negligible over the entire \PT-range due to the background-free identification based on the TPC d$E$/d$x$. 
Feed-down from weakly decaying hyper-tritons ($^{3}_{\Lambda}{\mathrm H}$) is negligible for deuterons~\cite{Adam:2015yta,nuclei}. Since only about 4-8\% of all $^{3}_{\Lambda}{\mathrm H}$ decaying into \he{} pass the track selection criteria for primary \he{}, the remaining contamination has not been subtracted and the uncertainty related to it was further investigated by a variation of the \dcaxy{}-cut in data and a final uncertainty of 5\% is assigned.
The influence of uncertainties in the material budget on the reconstruction efficiency has been studied by simulating events varying the amount of material by $\pm $10\%. The estimates of the uncertainties related to the tracking and ITS-TPC matching are based on a variation of the track cuts and are found to be approximately 5\%.
The uncertainties related to tracking, transport code, material budget and TPC-TOF matching are fully correlated across different multiplicity intervals.

\begin{table}[hb]
\caption{Main sources of systematic uncertainties for deuterons and \he{} as well as their anti-particles for low and high \PT{}.}
  \centering
  \begin{tabular}[hb]{l|cccc|cccc}
    \hline
    \hline  
	&&&\\[-1.0em]
    & \multicolumn{2}{c}{d} &\multicolumn{2}{c}{\ad} & \multicolumn{2}{c}{\he} &\multicolumn{2}{c}{\ahe}   \\
       \hline
    \PT{} (\gevc)                     & 0.9       & 2.9      & 0.9    & 2.9       & 2.2    & 5.0    & 1.8    & 5.0 \\
    \hline
    Tracking (ITS-TPC matching)                           & 5\%     & 5\%     & 5\%    & 5\%     & 6\%     & 4\%    & 6\%    & 4\% \\
    Secondaries material       & 1\%    & negl.   & negl.  & negl.   & 20\%   & 1\%    & negl.   & negl. \\
    Secondaries weak decay & negl.    & negl.   & negl.  & negl.   & 5\%     & negl.    &  5\%  & negl. \\
    Material budget                & 3\%     & 3\%     & 3\%    & 3\%     & 3\%     & 1\%    & 3\%    & 1\% \\
    Particle identification        & 1\%     & 3\%   & 1\%    & 3\%   & 3\%     & 3\%    & 3\%    & 3\%  \\
    Transport code         & 3\%     & 3\%     & 3\%   & 3\%  & 6\%     & 6\%    & 18\%  & 11\%\\
    TPC-TOF matching         & 3\%     & 3\%     & 5\%   & 5\%  & -     & -    & -  & -\\
    \hline
    Total			            & 7\%   & 8\%  & 8\% & 9\%    &  23\%  & 8\%   & 20\%   & 12\% \\
    \hline
    \hline
   \end{tabular}
   \label{tab:systematics}
\end{table}

\section{Results and Discussion}

\subsection{Spectra and yields}

\begin{figure}[htb]
	\begin{center}
	      \includegraphics[width = 0.49\textwidth]{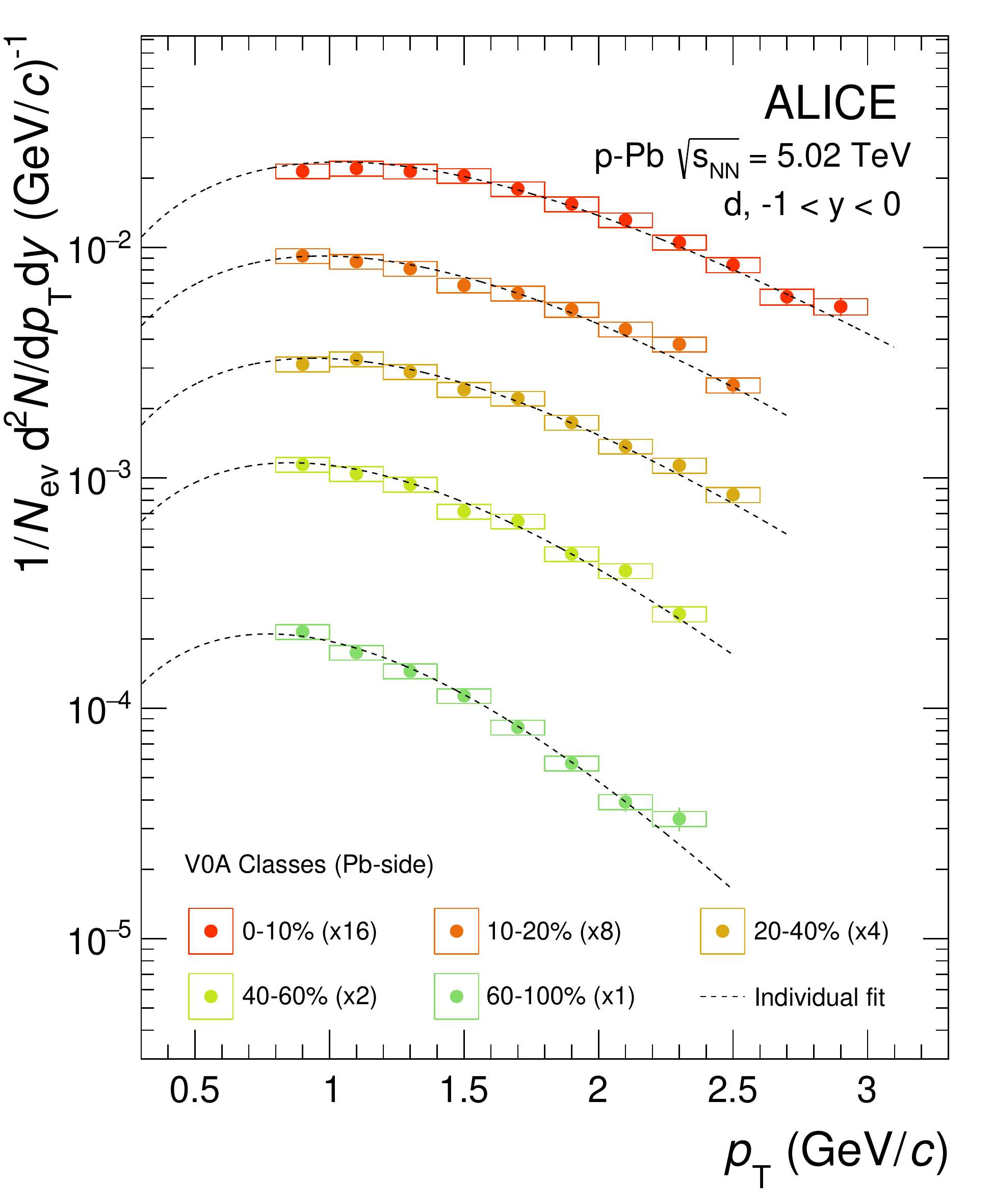} 
	      \includegraphics[width = 0.49\textwidth]{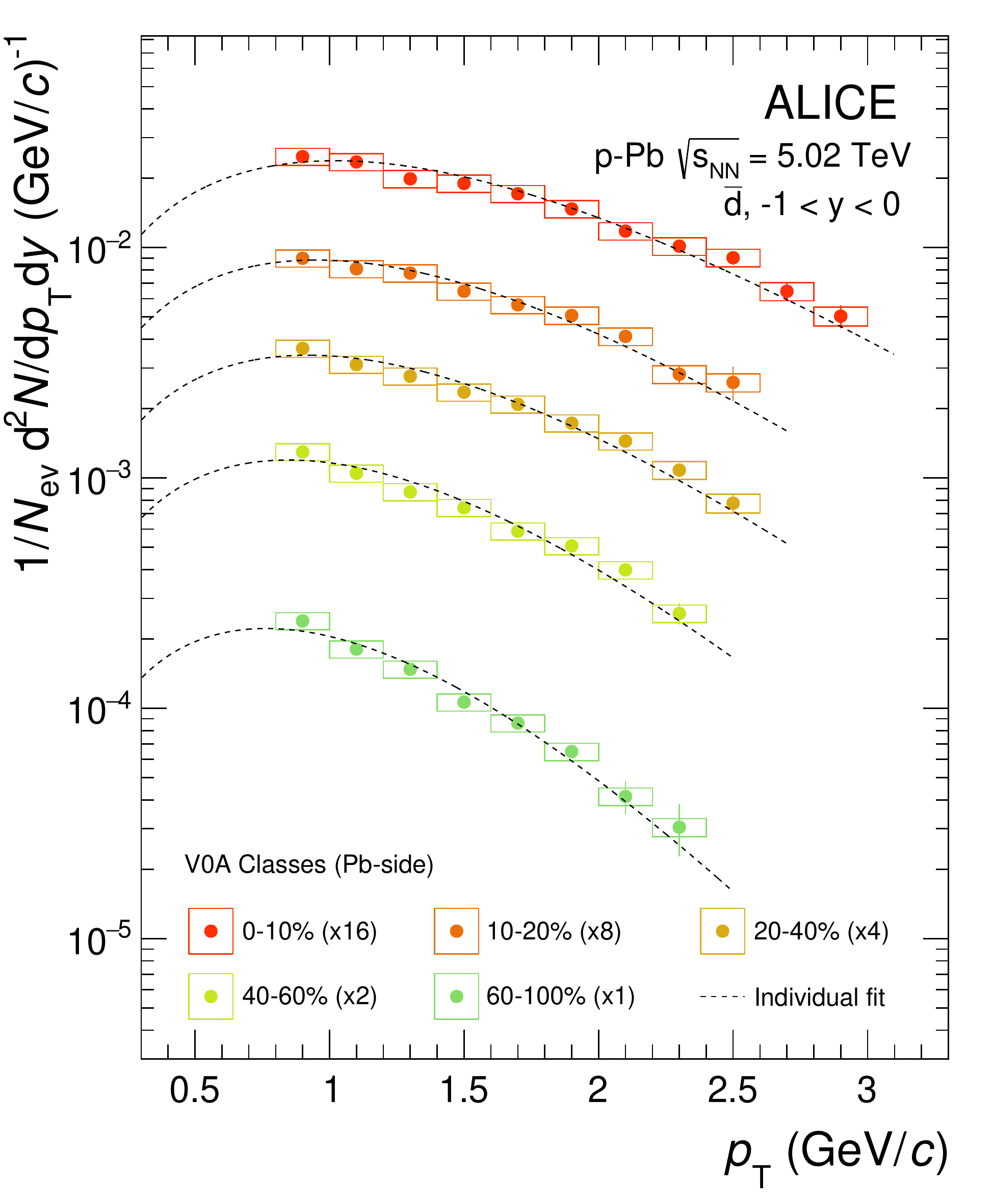}
	  \caption{Transverse momentum distributions of deuterons (left) and anti-deuterons (right) for various multiplicity classes. The multiplicity class definition is based on the signal amplitude observed in the V0A detector located on the Pb-side. The vertical bars represent the statistical errors,
	  the empty boxes show the systematic uncertainty. The lines represent individual fits using a $m_{\rm T}$-exponential function. }
	  \label{fig:spectra}
	\end{center}
\end{figure} 
%
%

The transverse momentum spectra of deuterons and anti-deuterons in the rapidity range $-1~<~y~<~0$  are presented in Fig.~\ref{fig:spectra} for several multiplicity classes. The spectra show a hardening with increasing event multiplicity. This behaviour was already observed for lower mass particles in \pPb{} collisions \cite{Abelev:2013haa}. For the extraction of \avpT{} and \PT-integrated yields \dndy{},  the spectra are fitted individually using a $m_{\rm T}$-exponential function \cite{Hagedorn:1965st}. 

\begin{figure}[htbp]
	\begin{center}
		\includegraphics[width = 0.55\textwidth]{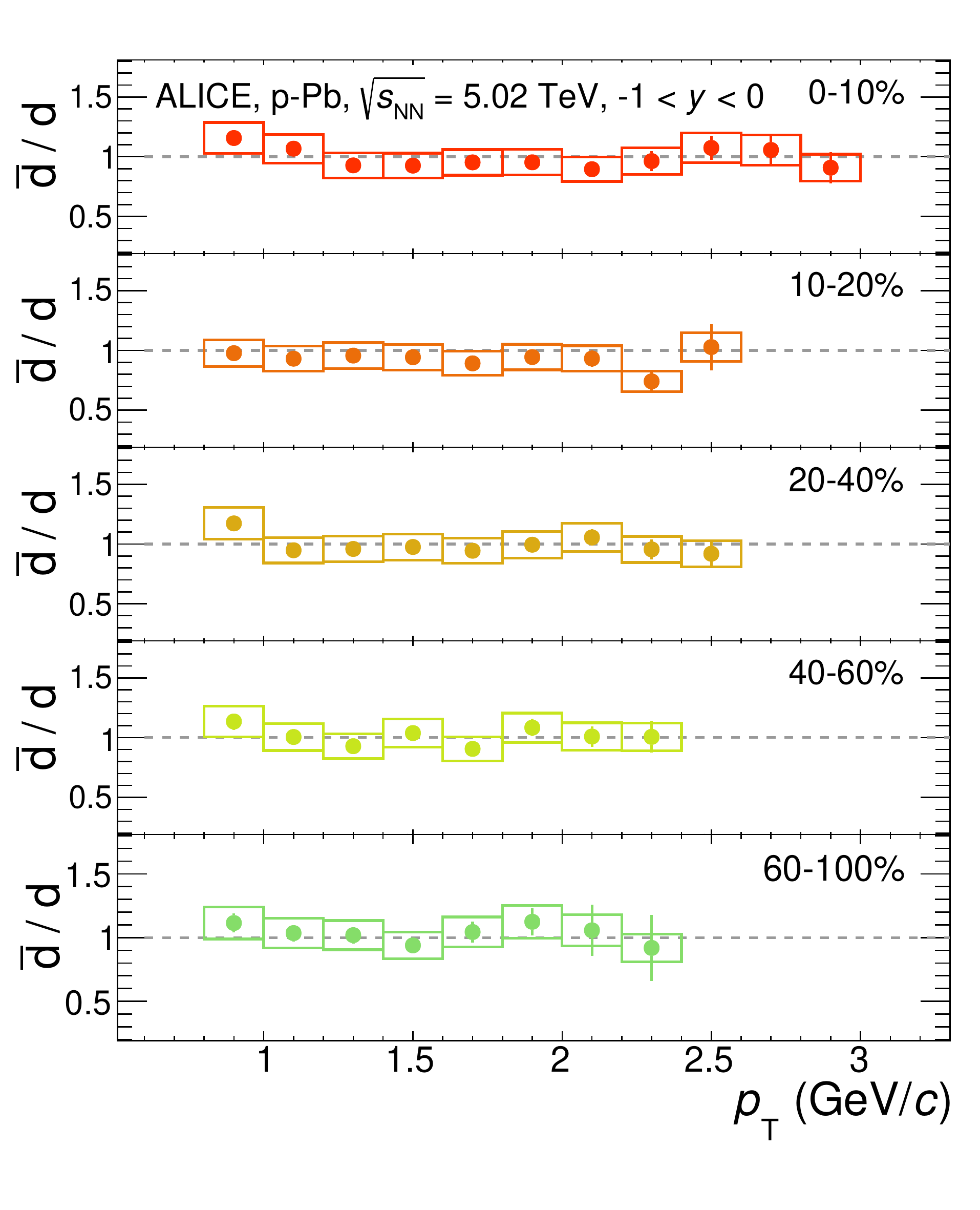} 
		\caption{Anti-deuteron to deuteron production ratio for the five multiplicity classes. All ratios are compatible with unity, indicated as a dashed grey line. 
		The vertical bars represent the statistical errors while the empty boxes show the total systematic uncertainty.
		}
		\label{fig:ratios}
	\end{center}
\end{figure}

The values obtained for \dndy{} for (anti-)deuterons are summarized in Table~\ref{tab:results}. They have been calculated by summing up the \PT-differential yield in the region where the spectrum is measured and by integrating the fit result in the unmeasured region at low and high transverse momenta. 
While the fraction of the extrapolated yield at high \PT{} is negligible, the fraction at low \PT{} ranges from 23\% at high to 38\% at low multiplicities. 
The uncertainty introduced by this extrapolation is estimated by comparing the result obtained with the $m_{\rm T}$-exponential fit to fit results from several alternative functional forms (Boltzmann, Blast-wave~\cite{Schnedermann:1993ws}, and \PT-exponential).

\begin{table}[htbp]
	\caption[Integrated]{Integrated yields \dndy{} of (anti-)deuterons. The first value is the statistical and the second is the total systematic uncertainty which includes both the systematic uncertainty on the measured spectra and the uncertainty of the extrapolation to low and high \PT.} 

	\centering
	{\small
	\begin{tabular}{ccc}
	
		\hline
		\hline
		&&\\[-0.7em]
	Multiplicity classes &  \dndy~(d) &  \dndy~(\ad) \\
		\hline
		&&\\[-0.7em]
0-10\%   & $(2.86\pm0.03\pm0.30)\times10^{-3} $ & $(2.83\pm0.03\pm0.35)\times10^{-3}$ \\
10-20\%  & $(2.08\pm0.02\pm0.22)\times10^{-3} $ & $(1.94\pm0.03\pm0.24)\times10^{-3}$ \\
20-40\%  & $(1.43\pm0.01\pm0.15)\times10^{-3} $ & $(1.43\pm0.02\pm0.17)\times10^{-3}$ \\
40-60\%  & $(8.93\pm0.08\pm0.93)\times10^{-4} $ & $(9.06\pm0.15\pm1.09)\times10^{-4}$ \\
60-100\% & $(2.89\pm0.05\pm0.30)\times10^{-4} $ & $(3.02\pm0.07\pm0.36)\times10^{-4}$ \\	&&\\[-0.7em]
	\hline
	\hline
 \end{tabular}
}
 \label{tab:results}
\end{table}

Figure~\ref{fig:ratios} shows the $\mathrm{\overline{d}}/\rm d$ ratios as a function of \PT{} for all multiplicity intervals. The ratios are found to be consistent with unity within uncertainties. This behaviour is expected, since thermal and coalescence models predict that the $\mathrm{\overline{d}}/\rm d$ ratio is given by $(\pbarp)^2$ (see for instance~\cite{Cleymans:2011pe}) and the $\pbarp$ ratio measured in \pPb{} collisions is consistent with unity for all multiplicity intervals~\cite{Abelev:2013haa}.

\begin{figure}[htbp]
	\begin{center}

		\includegraphics[width = 0.8\textwidth]{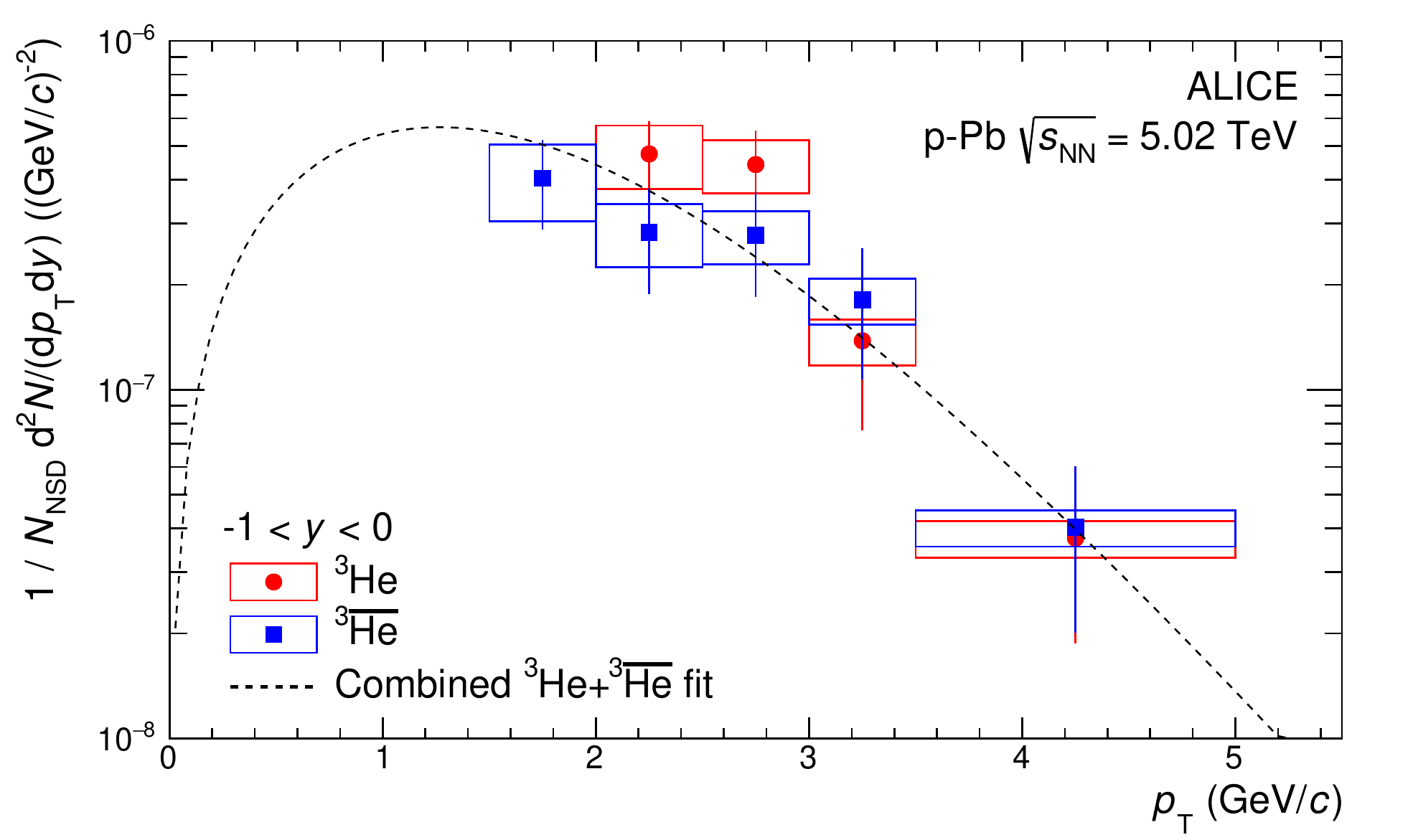}
		\caption{Transverse momentum distribution of \he{} and $^{3}\overline{\rm He}$ for all NSD collisions ($N_{\mathrm{NSD}}$). 
				The vertical bars represent the statistical errors while the empty boxes show the total systematic uncertainty.
		The line represents a $\chi^2$  fit with a $m_{\rm T}$-exponential function (see text for details).}
		\label{fig:spectrum-3}
	\end{center}
\end{figure}

The rare production of $A>2$ nuclei only allows the extraction of minimum-bias spectra for \he{} and \ahe{} with the available statistics and thus the result is normalised to all non-single diffractive (NSD) events. In total, 40 \ahe{} nuclei are observed, while about 29400 tracks from \ad{} are reconstructed in the same data sample. The corresponding spectra are shown in Fig.~\ref{fig:spectrum-3} together with a $m_{\rm T}$-exponential fit which is used for the extraction of the \dndy{} and \avpT{} of the spectra.  The fit is performed such that the residuals to both the \he{} and \ahe{} spectrum are minimised simultaneously. The fraction of the extrapolated yield corresponds to about 58\%. The uncertainty introduced by this extrapolation is also estimated by comparing the result obtained with the $m_{\rm T}$-exponential fit to fit results from several alternative functional forms (Boltzmann, Blast-wave~\cite{Schnedermann:1993ws}, and \PT-exponential). A \PT-integrated yield of \dndy~=~$(1.36 \pm 0.16 (\rm stat) \pm 0.52 (\rm syst))\times 10^{-6}$ and an average transverse momentum of  \avpT{}~=~$(1.78 \pm 0.11 (\rm stat) \pm 0.77 (\rm syst))$~\gevc{} are obtained.

%
%

\begin{figure}[htbp]
\begin{center}
\includegraphics[trim=0cm 0cm 0cm 0cm, clip=true, width = 0.7\textwidth]{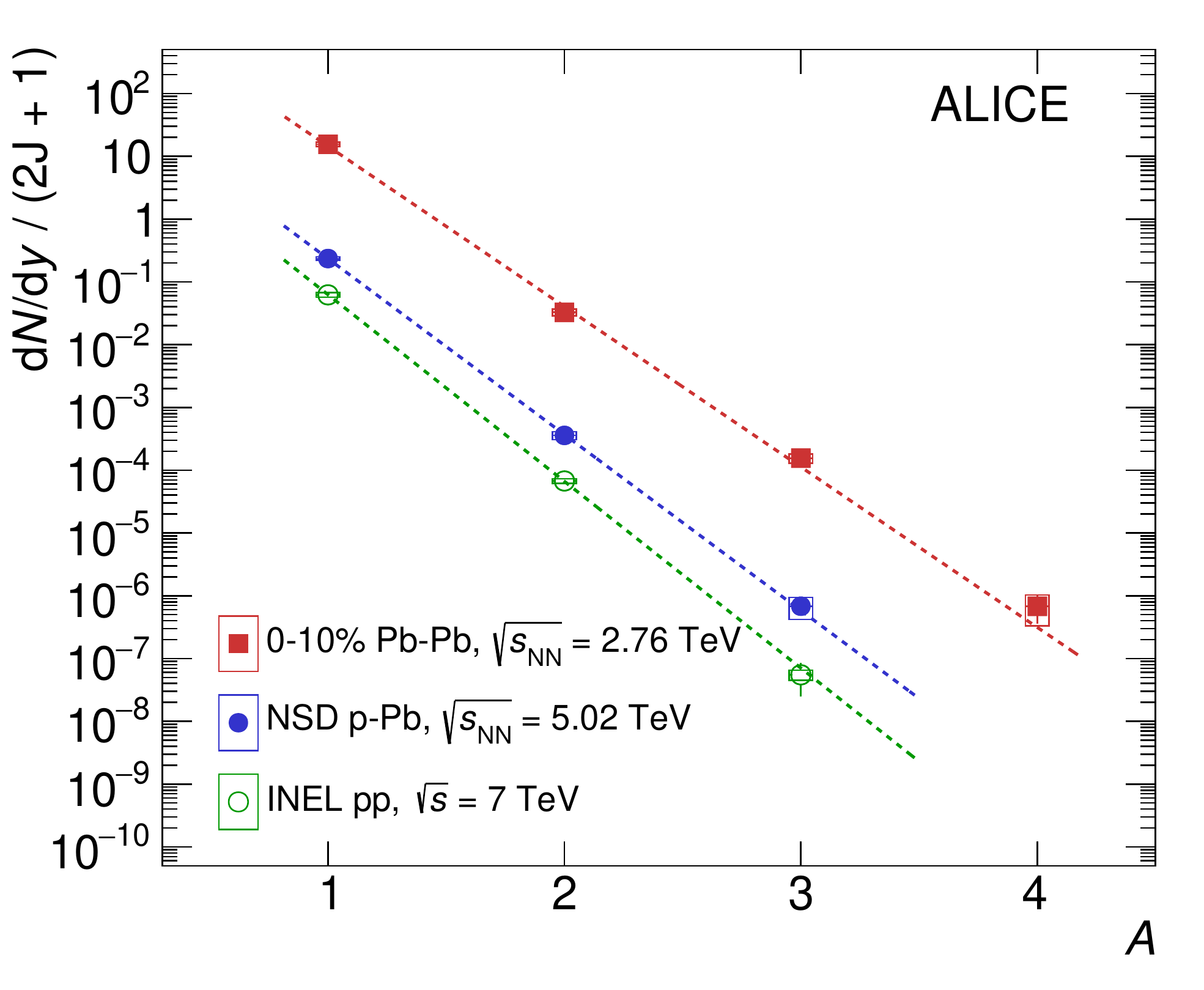} 	
		\caption{Production yield $\ensuremath{\mathrm{d}N/\mathrm{d}y}$ normalised by the spin degeneracy as a function of the mass number for inelastic pp collisions, minimum-bias p-Pb and central Pb-Pb collisions \cite{Acharya:2017bso,Acharya:2017fvb,Abelev:2013vea,Abelev:2013haa,Adam:2015qaa}. The empty boxes represent the total systematic uncertainty while the statistical errors are shown by the vertical bars. The lines represent fits with an exponential function.}
		\label{fig:yield_mass}
	\end{center}
\end{figure}

The yields of p, d and \he{} for NSD p--Pb events and normalised to their spin degeneracy are shown in Fig.~\ref{fig:yield_mass} as a function of the mass number $A$ together with results for inelastic pp collisions and central Pb-Pb collisions. An exponential decrease with increasing $A$ is observed in all cases, yet with different slopes. 
The penalty factor, i.e. the reduction of the yield for each additional nucleon, is obtained from a fit to the data and a value of 
$635 \pm 90$ in p-Pb collisions is found which is significantly 
larger than the factor of $359 \pm 41$ 
which was observed for central \PbPb{} collisions~\cite{nuclei}. 
The penalty factor obtained for the inelastic pp collisions  \cite{Acharya:2017fvb} is found to be $942 \pm 107$.
Such an exponential decrease of the \mbox{(anti-)}nuclei yield with mass number has also been observed at lower incident energies in heavy-ion \cite{Barrette:1994tw, BraunMunzinger:1994iq, Arsenescu:2003eg, Agakishiev:2011ib} as well as in p--A collisions \cite{Antipov:1970uc}.

%
%

\subsection{Coalescence parameter}

\begin{figure}[htbp]	
  \begin{center}
    \includegraphics[width = 0.6\textwidth]{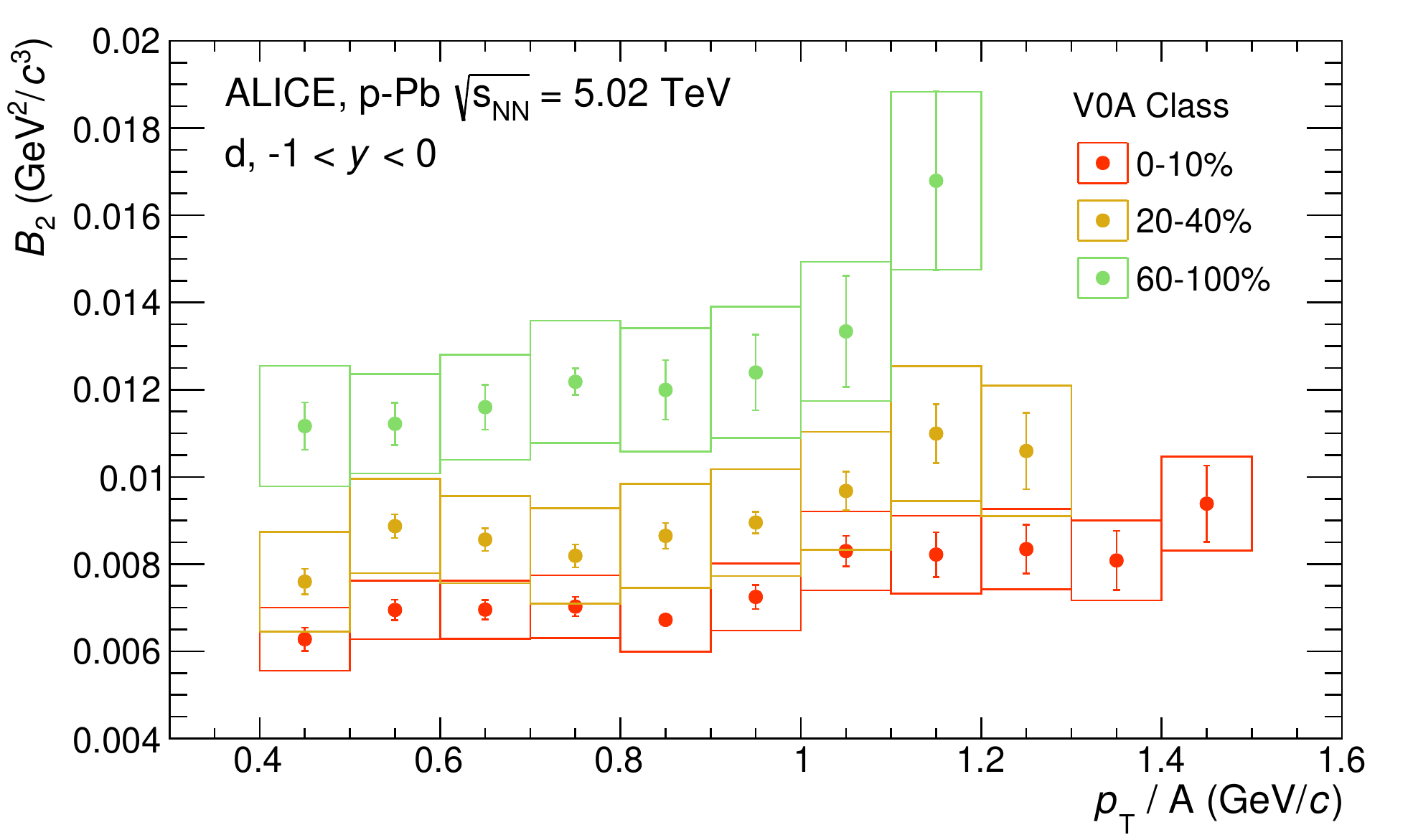} 
    \caption{Coalescence parameter $B_2$ as a function of \PT{} for different V0A multiplicity classes. 
    		The vertical lines represent the statistical errors and the empty boxes show the total systematic uncertainty.
	}
    \label{fig:b2}
  \end{center}
\end{figure}

In the traditional coalescence model, deuterons and other light nuclei are formed by protons and neutrons, which are close in phase space. In this picture, the deuteron momentum spectra are related to those of its constituent nucleons via~\cite{Butler:1963pp,Scheibl:1998tk} 

\begin{equation}
E_{\rm d} \, \frac{{\rm d^3} N_{\rm d}}{{\rm d} p_{\rm d}^3} =  B_2 \, \left(
E_{\rm p} \, \frac{{\rm d^3} N_{\rm p}}{{\rm d} p_{\rm p}^3}
\right)^2,
\label{coal}
\end{equation}

where the momentum of the deuteron is given by $p_{\mathrm{d}} = 2 p_{\mathrm{p}}$. Since the neutron spectra are experimentally not accessible, they are approximated by the proton spectra. The value of \Btwo{} is computed as a function of event multiplicity and transverse momentum as the ratio between the deuteron yield measured at $p_{\mathrm{T}}=p_{\mathrm{T,d}}$ and the square of the proton yield at $p_{\mathrm{T,p}} = 0.5p_{\mathrm{T,d}}$. The obtained \Btwo{}-values are shown in Fig.~\ref{fig:b2}. In its simplest implementation, the coalescence model for uncorrelated particle emission from a point-like source predicts that the observed \Btwo{}-values are independent of \PT{} and of event multiplicity (called "simple coalescence" in the following). 
Within uncertainties and given the current width of the multiplicity classes, the observed \PT{} dependence is still compatible with the expected flat behaviour (for a detailed discussion see \cite{Acharya:2019rgc}).
Moreover, a decrease of the measured \Btwo{} parameter with increasing event multiplicity for a fixed \PT{} is observed. This effect is even more pronounced in \PbPb{} collisions \cite{nuclei} and a possible explanation is an increasing source volume, which can effectively reduce the coalescence probability~\cite{Scheibl:1998tk,Blum:2017qnn}.

%
%

\subsection{Mean transverse momenta}

\begin{figure}[htbp]
  \begin{center}
     \includegraphics[trim=0cm 0cm 0cm 0cm, clip=true, width = 1.0\textwidth]{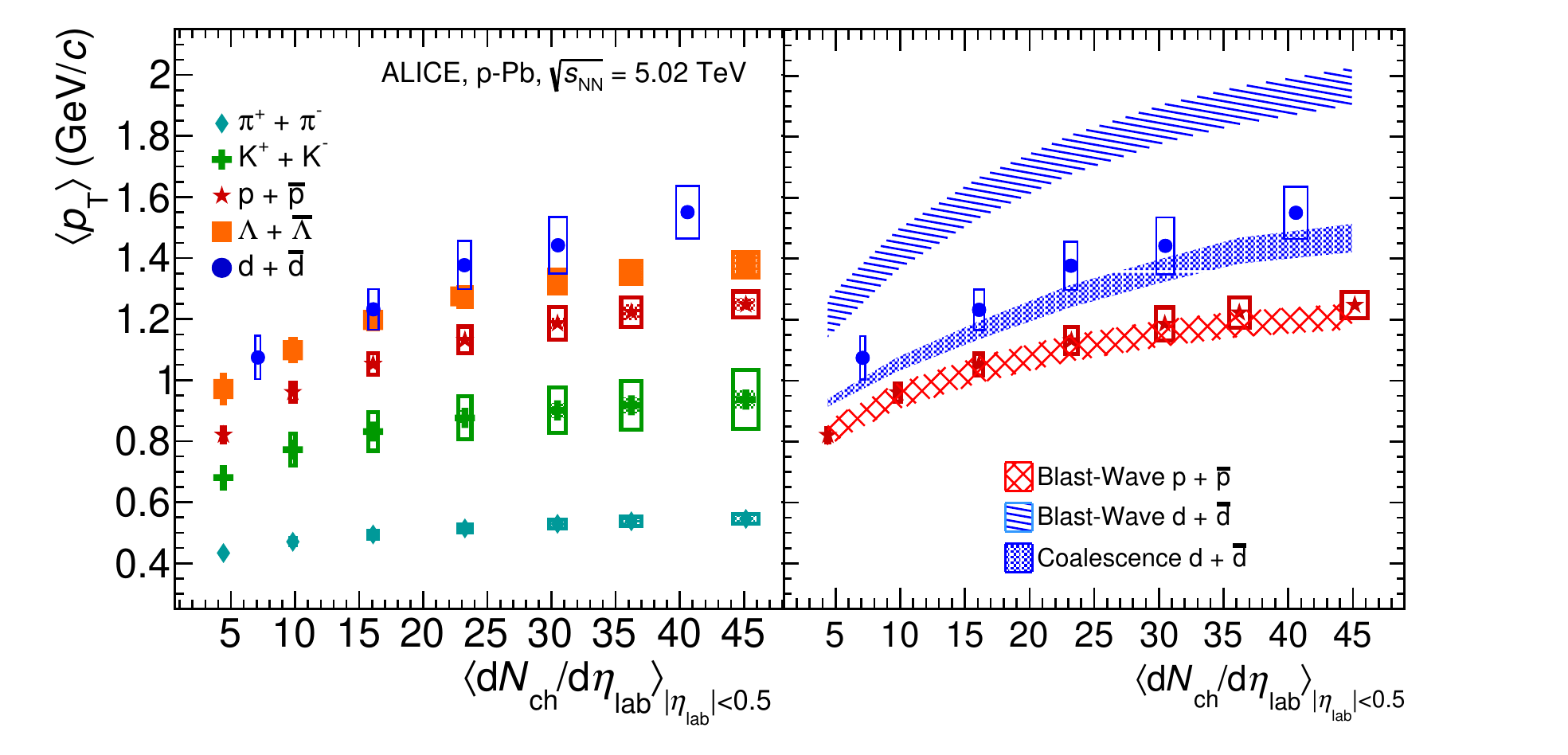}
    \caption{Mean \PT{} of various particle species as a function of the mean charged-particle density at mid-rapidity for different V0A multiplicity classes. The empty boxes show the total systematic uncertainty while the shaded boxes indicate the contribution which is uncorrelated across multiplicity intervals (left). 
    Comparison of \avpT{} of protons and deuterons with the simple coalescence and the Blast-Wave model  expectations. The shaded areas show the expected \avpT{} for deuterons from a simple coalescence model assuming a \PT-independent \Btwo{} as well as the calculated \avpT{} for protons and deuterons from the Blast-Wave model~\cite{Schnedermann:1993ws} using the kinetic freeze-out parameters for pions, kaons, protons and $\Lambda$ from~\cite{Abelev:2013haa} (right).}
    \label{fig:meanpt}
  \end{center}
\end{figure}

In Fig.~\ref{fig:meanpt} (left), the mean values of the transverse momenta of deuterons are compared with the corresponding results for 
$\pi^{\pm}$, K$^{\pm}$, p($\rm \bar{p}$), and $\Lambda$($\overline{\Lambda}$)~\cite{Abelev:2013haa}. 
As for all other particles, the \avpT{} of deuterons shows an increase with increasing event multiplicity, which reflects the observed hardening of the spectra. However, it is striking that deuterons violate the mass ordering which was observed for non-composite particles~\cite{Abelev:2013haa,Adam:2016bpr}: despite their much larger mass, the \avpT{} values are similar to those of $\Lambda$($\overline{\Lambda}$) and only slightly higher than those of p($\rm \bar{p}$).

Note that simple coalescence models give a significantly different prediction for the \avpT{} of deuterons with respect to hydrodynamical models.
This can be best illustrated with two simplifying requirements which are approximately fulfilled in data. Firstly, the coalescence parameter is assumed flat in \PT{} and secondly the proton spectrum can be described by an exponential shape, i.e. $C \exp(-\PT/T)$ with two parameters $C$ and $T$. In this case, the shape of the deuteron spectrum can be analytically calculated based on the definition of \Btwo{}. Due to the self-similarity feature of the exponential function, $ \left(\exp(x/a)\right)^{a} = \exp(x) $, the spectral shape of the proton and the deuteron are then found to be identical:

\begin{equation}
 \frac{1}{2\pi p_{\rm T}^{d}}  \frac{{\rm d}^{2}N^{d}}{{\rm d}y \,{\rm d}p_{\rm T}^{d}} = B_{2} 
\Bigl(  \frac{1}{ 2\pi p_{\rm T}^{p}}  \frac{{\rm d}^{2}N^{p} }{ {\rm d}y \,{\rm d}p_{\rm T}^{p}} \Bigr)^{2} =
B_{2} \Bigl(  C \exp(-\frac{p_{\rm T}^{p} }{ T}) \Bigr)^{2} = B_{2} \Bigl(  C \exp(-\frac{p_{\rm T}^{d} }{ 2T}) \Bigr)^{2} = B_{2}~C^{2}\exp(-\frac{p_{\rm T}^{d} }{ T})
\;.
\label{eq.:B2expression}
\end{equation}

Thus, the same \avpT{} for both particles is expected and the behaviour observed in p--Pb collisions is well described by simple coalescence models. This finding can be even further substantiated by directly calculating the \avpT{} of deuterons assuming a constant value of \Btwo{} and using the measured proton spectrum as input. As shown in Fig.~\ref{fig:meanpt} (right), in this case, a good agreement with the data is found considering that a large fraction of the systematic uncertainty is correlated among different multiplicity bins. The Blast-Wave model~\cite{Schnedermann:1993ws} fails to describe the \avpT{} values for deuterons using the common kinetic freeze-out parameters from~\cite{Abelev:2013haa}, which describe simultaneously the spectra of pions, kaons, and protons.

%
%

\subsection{Deuteron-over-proton ratio}

\begin{figure}[htbp]
  \begin{center}
    \includegraphics[trim=0cm 0cm 0cm 0cm, clip=true, width = 0.6\textwidth]{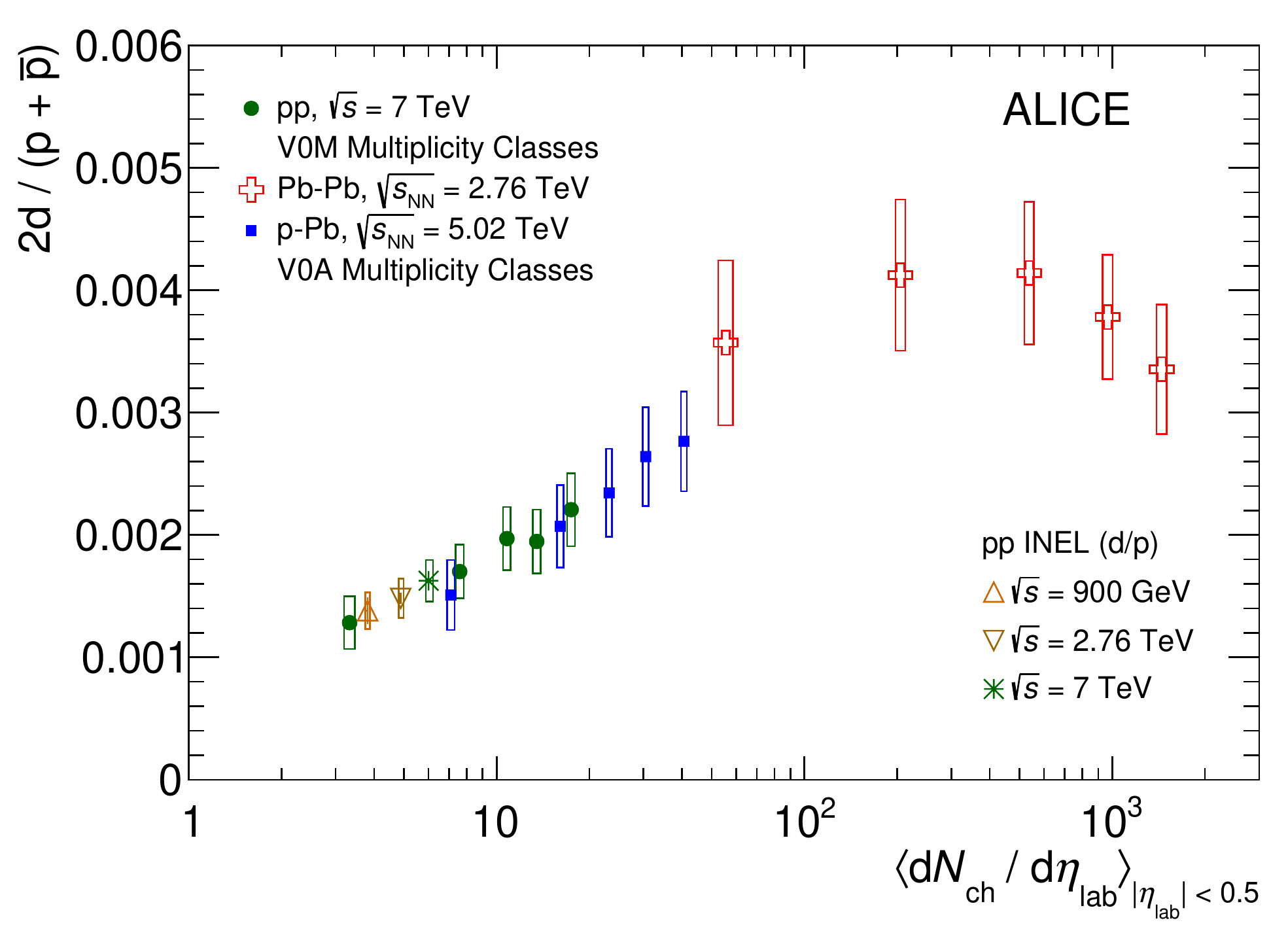} 
    \caption{Deuteron-over-proton ratio as a function of charged-particle multiplicity at mid-rapidity for \pp, \pPb{} and \PbPb{} collisions \cite{Acharya:2019rgc,nuclei,Acharya:2017fvb}. The empty boxes show the systematic uncertainty while the vertical lines represent the statistical uncertainty.}
    \label{fig:doverp}
  \end{center}
\end{figure}

The deuteron-over-proton ratio is shown in Fig.~\ref{fig:doverp} for three collision systems as a function of the charged-particle density at mid-rapidity. In \PbPb{} collisions it has been observed that the d/p ratio does not vary 
with centrality within uncertainties 
(red symbols). Such a trend is consistent with a thermal-statistical approach and the magnitude of the measured values agree with freeze-out temperatures in the range of 150-160~MeV \cite{nuclei}. 
The d/p ratio obtained in inelastic \pp{} collisions increases with multiplicity~\cite{Acharya:2019rgc}.
The results in \pPb{} collisions bridge the two measurements in terms of multiplicity and system size and show an increase of the d/p ratio with multiplicity. Here, the low (high) multiplicity value is compatible with the result from pp (\PbPb) collisions. 
Note that the experimental significance of this enhancement is further substantiated by considering only the part of the systematic uncertainty which is uncorrelated across multiplicity intervals.

%
%

A similar rise with multiplicity is observed for the ratios of the yields of multi-strange particles to that of pions in \pPb{} collisions~\cite{multistrange-p-Pb}. In this case the canonical suppression due to exact strangeness conservation in smaller systems gives a qualitative explanation \cite{Acharya:2018orn}. An interpretation of the d/p ratio within thermal models is difficult, since the measured \ppi{} ratio in these three systems is about the same \cite{Abelev:2013haa}. Therefore, the available parameter space for a change in the freeze-out temperature or a suppression due to exact conservation of baryon number is limited \cite{Vovchenko:2018fiy}.
Coalescence models are able to explain such an observation. The probability of forming a deuteron increases with the nucleon density and thus also with the charged-particle density. The results from pp and \pPb{} collisions at low charged-particle density fit with this concept.

\section{Conclusions}

The production of deuterons and \he{} and their antiparticles in \pPb{} collisions at \snn{} = 5.02 TeV  has been studied at mid-rapidity. The results on deuteron production in \pPb{} collisions exhibit a continuous evolution with multiplicity  between \pp{} and \PbPb{} collisions. 
The production of complex nuclei shows an exponential decrease with mass (number). The penalty factor (decrease of yield for each additional nucleon) is larger than the one observed in central Pb--Pb collisions and smaller than the one measured in pp collisions. The transverse momentum distributions of deuterons become harder with increasing multiplicity. Two intriguing observations that have been recently reported by ALICE \cite{Acharya:2019rgc} in high multiplicity pp collisions are confirmed in the present paper.
Firstly, the \avpT{} values of deuterons are comparable to those of the much lighter $\Lambda$ baryons and thus do not follow a mass ordering.
This behaviour is observed for all multiplicity intervals and it is in contrast to the expectation from simple hydrodynamical models. 
These observations made in \pPb{} collisions support a coalescence mechanism, while in \PbPb{} collisions the deuteron seems to follow the collective expansion of the fireball. Secondly, the d/p ratio rises strongly with multiplicity, while this ratio remains approximately constant as a function of multiplicity in \PbPb{} collisions, where its value agrees with thermal-model predictions. 

\newenvironment{acknowledgement}{\relax}{\relax}
\begin{acknowledgement}
\section*{Acknowledgements}

The ALICE Collaboration would like to thank all its engineers and technicians for their invaluable contributions to the construction of the experiment and the CERN accelerator teams for the outstanding performance of the LHC complex.
The ALICE Collaboration gratefully acknowledges the resources and support provided by all Grid centres and the Worldwide LHC Computing Grid (WLCG) collaboration.
The ALICE Collaboration acknowledges the following funding agencies for their support in building and running the ALICE detector:
A. I. Alikhanyan National Science Laboratory (Yerevan Physics Institute) Foundation (ANSL), State Committee of Science and World Federation of Scientists (WFS), Armenia;
Austrian Academy of Sciences, Austrian Science Fund (FWF): [M 2467-N36] and Nationalstiftung f\"{u}r Forschung, Technologie und Entwicklung, Austria;
Ministry of Communications and High Technologies, National Nuclear Research Center, Azerbaijan;
Conselho Nacional de Desenvolvimento Cient\'{\i}fico e Tecnol\'{o}gico (CNPq), Universidade Federal do Rio Grande do Sul (UFRGS), Financiadora de Estudos e Projetos (Finep) and Funda\c{c}\~{a}o de Amparo \`{a} Pesquisa do Estado de S\~{a}o Paulo (FAPESP), Brazil;
Ministry of Science \& Technology of China (MSTC), National Natural Science Foundation of China (NSFC) and Ministry of Education of China (MOEC) , China;
Croatian Science Foundation and Ministry of Science and Education, Croatia;
Centro de Aplicaciones Tecnol\'{o}gicas y Desarrollo Nuclear (CEADEN), Cubaenerg\'{\i}a, Cuba;
Ministry of Education, Youth and Sports of the Czech Republic, Czech Republic;
The Danish Council for Independent Research | Natural Sciences, the Carlsberg Foundation and Danish National Research Foundation (DNRF), Denmark;
Helsinki Institute of Physics (HIP), Finland;
Commissariat \`{a} l'Energie Atomique (CEA), Institut National de Physique Nucl\'{e}aire et de Physique des Particules (IN2P3) and Centre National de la Recherche Scientifique (CNRS) and R\'{e}gion des  Pays de la Loire, France;
Bundesministerium f\"{u}r Bildung und Forschung (BMBF) and GSI Helmholtzzentrum f\"{u}r Schwerionenforschung GmbH, Germany;
General Secretariat for Research and Technology, Ministry of Education, Research and Religions, Greece;
National Research, Development and Innovation Office, Hungary;
Department of Atomic Energy Government of India (DAE), Department of Science and Technology, Government of India (DST), University Grants Commission, Government of India (UGC) and Council of Scientific and Industrial Research (CSIR), India;
Indonesian Institute of Science, Indonesia;
Centro Fermi - Museo Storico della Fisica e Centro Studi e Ricerche Enrico Fermi and Istituto Nazionale di Fisica Nucleare (INFN), Italy;
Institute for Innovative Science and Technology , Nagasaki Institute of Applied Science (IIST), Japan Society for the Promotion of Science (JSPS) KAKENHI and Japanese Ministry of Education, Culture, Sports, Science and Technology (MEXT), Japan;
Consejo Nacional de Ciencia (CONACYT) y Tecnolog\'{i}a, through Fondo de Cooperaci\'{o}n Internacional en Ciencia y Tecnolog\'{i}a (FONCICYT) and Direcci\'{o}n General de Asuntos del Personal Academico (DGAPA), Mexico;
Nederlandse Organisatie voor Wetenschappelijk Onderzoek (NWO), Netherlands;
The Research Council of Norway, Norway;
Commission on Science and Technology for Sustainable Development in the South (COMSATS), Pakistan;
Pontificia Universidad Cat\'{o}lica del Per\'{u}, Peru;
Ministry of Science and Higher Education and National Science Centre, Poland;
Korea Institute of Science and Technology Information and National Research Foundation of Korea (NRF), Republic of Korea;
Ministry of Education and Scientific Research, Institute of Atomic Physics and Ministry of Research and Innovation and Institute of Atomic Physics, Romania;
Joint Institute for Nuclear Research (JINR), Ministry of Education and Science of the Russian Federation, National Research Centre Kurchatov Institute, Russian Science Foundation and Russian Foundation for Basic Research, Russia;
Ministry of Education, Science, Research and Sport of the Slovak Republic, Slovakia;
National Research Foundation of South Africa, South Africa;
Swedish Research Council (VR) and Knut \& Alice Wallenberg Foundation (KAW), Sweden;
European Organization for Nuclear Research, Switzerland;
National Science and Technology Development Agency (NSDTA), Suranaree University of Technology (SUT) and Office of the Higher Education Commission under NRU project of Thailand, Thailand;
Turkish Atomic Energy Agency (TAEK), Turkey;
National Academy of  Sciences of Ukraine, Ukraine;
Science and Technology Facilities Council (STFC), United Kingdom;
National Science Foundation of the United States of America (NSF) and United States Department of Energy, Office of Nuclear Physics (DOE NP), United States of America.    
\end{acknowledgement}

\bibliographystyle{utphys} 
\bibliography{main}  

\newpage
\appendix
\section{The ALICE Collaboration}
\label{app:collab}

\begingroup
\small
\begin{flushleft}
S.~Acharya\Irefn{org141}\And 
D.~Adamov\'{a}\Irefn{org93}\And 
S.P.~Adhya\Irefn{org141}\And 
A.~Adler\Irefn{org73}\And 
J.~Adolfsson\Irefn{org79}\And 
M.M.~Aggarwal\Irefn{org98}\And 
G.~Aglieri Rinella\Irefn{org34}\And 
M.~Agnello\Irefn{org31}\And 
N.~Agrawal\Irefn{org48}\textsuperscript{,}\Irefn{org10}\And 
Z.~Ahammed\Irefn{org141}
% ...
\And
\renewcommand\labelenumi{\textsuperscript{\theenumi}~}

\section*{Affiliation notes}
\renewcommand\theenumi{\roman{enumi}}
\begin{Authlist}
\item \Adef{org*}Deceased
\item \Adef{orgI}Dipartimento DET del Politecnico di Torino, Turin, Italy
\item \Adef{orgII}M.V. Lomonosov Moscow State University, D.V. Skobeltsyn Institute of Nuclear, Physics, Moscow, Russia
\item \Adef{orgIII}Department of Applied Physics, Aligarh Muslim University, Aligarh, India
\item \Adef{orgIV}Institute of Theoretical Physics, University of Wroclaw, Poland
\item \Adef{orgV}PINSTECH, Islamabad, Pakistan
\end{Authlist}

\section*{Collaboration Institutes}
\renewcommand\theenumi{\arabic{enumi}~}
\begin{Authlist}
\item \Idef{org1}A.I. Alikhanyan National Science Laboratory (Yerevan Physics Institute) Foundation, Yerevan, Armenia
\item \Idef{org2}Bogolyubov Institute for Theoretical Physics, National Academy of Sciences of Ukraine, Kiev, Ukraine
\item \Idef{org3}Bose Institute, Department of Physics  and Centre for Astroparticle Physics and Space Science (CAPSS), Kolkata, India
\item \Idef{org4}Budker Institute for Nuclear Physics, Novosibirsk, Russia
\item \Idef{org5}California Polytechnic State University, San Luis Obispo, California, United States
\item \Idef{org6}Central China Normal University, Wuhan, China
\item \Idef{org7}Centre de Calcul de l'IN2P3, Villeurbanne, Lyon, France
\item \Idef{org8}Centro de Aplicaciones Tecnol\'{o}gicas y Desarrollo Nuclear (CEADEN), Havana, Cuba
\item \Idef{org9}Centro de Investigaci\'{o}n y de Estudios Avanzados (CINVESTAV), Mexico City and M\'{e}rida, Mexico
\item \Idef{org10}Centro Fermi - Museo Storico della Fisica e Centro Studi e Ricerche ``Enrico Fermi', Rome, Italy
\item \Idef{org11}Chicago State University, Chicago, Illinois, United States
\item \Idef{org12}China Institute of Atomic Energy, Beijing, China
\item \Idef{org13}Chonbuk National University, Jeonju, Republic of Korea
\item \Idef{org14}Comenius University Bratislava, Faculty of Mathematics, Physics and Informatics, Bratislava, Slovakia
\item \Idef{org15}COMSATS University Islamabad, Islamabad, Pakistan
\item \Idef{org16}Creighton University, Omaha, Nebraska, United States
\item \Idef{org17}Department of Physics, Aligarh Muslim University, Aligarh, India
\item \Idef{org18}Department of Physics, Pusan National University, Pusan, Republic of Korea
\item \Idef{org19}Department of Physics, Sejong University, Seoul, Republic of Korea
\item \Idef{org20}Department of Physics, University of California, Berkeley, California, United States
\item \Idef{org21}Department of Physics, University of Oslo, Oslo, Norway
\item \Idef{org22}Department of Physics and Technology, University of Bergen, Bergen, Norway
\item \Idef{org23}Dipartimento di Fisica dell'Universit\`{a} 'La Sapienza' and Sezione INFN, Rome, Italy
\item \Idef{org24}Dipartimento di Fisica dell'Universit\`{a} and Sezione INFN, Cagliari, Italy
\item \Idef{org25}Dipartimento di Fisica dell'Universit\`{a} and Sezione INFN, Trieste, Italy
\item \Idef{org26}Dipartimento di Fisica dell'Universit\`{a} and Sezione INFN, Turin, Italy
\item \Idef{org27}Dipartimento di Fisica e Astronomia dell'Universit\`{a} and Sezione INFN, Bologna, Italy
\item \Idef{org28}Dipartimento di Fisica e Astronomia dell'Universit\`{a} and Sezione INFN, Catania, Italy
\item \Idef{org29}Dipartimento di Fisica e Astronomia dell'Universit\`{a} and Sezione INFN, Padova, Italy
\item \Idef{org30}Dipartimento di Fisica `E.R.~Caianiello' dell'Universit\`{a} and Gruppo Collegato INFN, Salerno, Italy
\item \Idef{org31}Dipartimento DISAT del Politecnico and Sezione INFN, Turin, Italy
\item \Idef{org32}Dipartimento di Scienze e Innovazione Tecnologica dell'Universit\`{a} del Piemonte Orientale and INFN Sezione di Torino, Alessandria, Italy
\item \Idef{org33}Dipartimento Interateneo di Fisica `M.~Merlin' and Sezione INFN, Bari, Italy
\item \Idef{org34}European Organization for Nuclear Research (CERN), Geneva, Switzerland
\item \Idef{org35}Faculty of Electrical Engineering, Mechanical Engineering and Naval Architecture, University of Split, Split, Croatia
\item \Idef{org36}Faculty of Engineering and Science, Western Norway University of Applied Sciences, Bergen, Norway
\item \Idef{org37}Faculty of Nuclear Sciences and Physical Engineering, Czech Technical University in Prague, Prague, Czech Republic
\item \Idef{org38}Faculty of Science, P.J.~\v{S}af\'{a}rik University, Ko\v{s}ice, Slovakia
\item \Idef{org39}Frankfurt Institute for Advanced Studies, Johann Wolfgang Goethe-Universit\"{a}t Frankfurt, Frankfurt, Germany
\item \Idef{org40}Gangneung-Wonju National University, Gangneung, Republic of Korea
\item \Idef{org41}Gauhati University, Department of Physics, Guwahati, India
\item \Idef{org42}Helmholtz-Institut f\"{u}r Strahlen- und Kernphysik, Rheinische Friedrich-Wilhelms-Universit\"{a}t Bonn, Bonn, Germany
\item \Idef{org43}Helsinki Institute of Physics (HIP), Helsinki, Finland
\item \Idef{org44}High Energy Physics Group,  Universidad Aut\'{o}noma de Puebla, Puebla, Mexico
\item \Idef{org45}Hiroshima University, Hiroshima, Japan
\item \Idef{org46}Hochschule Worms, Zentrum  f\"{u}r Technologietransfer und Telekommunikation (ZTT), Worms, Germany
\item \Idef{org47}Horia Hulubei National Institute of Physics and Nuclear Engineering, Bucharest, Romania
\item \Idef{org48}Indian Institute of Technology Bombay (IIT), Mumbai, India
\item \Idef{org49}Indian Institute of Technology Indore, Indore, India
\item \Idef{org50}Indonesian Institute of Sciences, Jakarta, Indonesia
\item \Idef{org51}INFN, Laboratori Nazionali di Frascati, Frascati, Italy
\item \Idef{org52}INFN, Sezione di Bari, Bari, Italy
\item \Idef{org53}INFN, Sezione di Bologna, Bologna, Italy
\item \Idef{org54}INFN, Sezione di Cagliari, Cagliari, Italy
\item \Idef{org55}INFN, Sezione di Catania, Catania, Italy
\item \Idef{org56}INFN, Sezione di Padova, Padova, Italy
\item \Idef{org57}INFN, Sezione di Roma, Rome, Italy
\item \Idef{org58}INFN, Sezione di Torino, Turin, Italy
\item \Idef{org59}INFN, Sezione di Trieste, Trieste, Italy
\item \Idef{org60}Inha University, Incheon, Republic of Korea
\item \Idef{org61}Institut de Physique Nucl\'{e}aire d'Orsay (IPNO), Institut National de Physique Nucl\'{e}aire et de Physique des Particules (IN2P3/CNRS), Universit\'{e} de Paris-Sud, Universit\'{e} Paris-Saclay, Orsay, France
\item \Idef{org62}Institute for Nuclear Research, Academy of Sciences, Moscow, Russia
\item \Idef{org63}Institute for Subatomic Physics, Utrecht University/Nikhef, Utrecht, Netherlands
\item \Idef{org64}Institute of Experimental Physics, Slovak Academy of Sciences, Ko\v{s}ice, Slovakia
\item \Idef{org65}Institute of Physics, Homi Bhabha National Institute, Bhubaneswar, India
\item \Idef{org66}Institute of Physics of the Czech Academy of Sciences, Prague, Czech Republic
\item \Idef{org67}Institute of Space Science (ISS), Bucharest, Romania
\item \Idef{org68}Institut f\"{u}r Kernphysik, Johann Wolfgang Goethe-Universit\"{a}t Frankfurt, Frankfurt, Germany
\item \Idef{org69}Instituto de Ciencias Nucleares, Universidad Nacional Aut\'{o}noma de M\'{e}xico, Mexico City, Mexico
\item \Idef{org70}Instituto de F\'{i}sica, Universidade Federal do Rio Grande do Sul (UFRGS), Porto Alegre, Brazil
\item \Idef{org71}Instituto de F\'{\i}sica, Universidad Nacional Aut\'{o}noma de M\'{e}xico, Mexico City, Mexico
\item \Idef{org72}iThemba LABS, National Research Foundation, Somerset West, South Africa
\item \Idef{org73}Johann-Wolfgang-Goethe Universit\"{a}t Frankfurt Institut f\"{u}r Informatik, Fachbereich Informatik und Mathematik, Frankfurt, Germany
\item \Idef{org74}Joint Institute for Nuclear Research (JINR), Dubna, Russia
\item \Idef{org75}Korea Institute of Science and Technology Information, Daejeon, Republic of Korea
\item \Idef{org76}KTO Karatay University, Konya, Turkey
\item \Idef{org77}Laboratoire de Physique Subatomique et de Cosmologie, Universit\'{e} Grenoble-Alpes, CNRS-IN2P3, Grenoble, France
\item \Idef{org78}Lawrence Berkeley National Laboratory, Berkeley, California, United States
\item \Idef{org79}Lund University Department of Physics, Division of Particle Physics, Lund, Sweden
\item \Idef{org80}Nagasaki Institute of Applied Science, Nagasaki, Japan
\item \Idef{org81}Nara Women{'}s University (NWU), Nara, Japan
\item \Idef{org82}National and Kapodistrian University of Athens, School of Science, Department of Physics , Athens, Greece
\item \Idef{org83}National Centre for Nuclear Research, Warsaw, Poland
\item \Idef{org84}National Institute of Science Education and Research, Homi Bhabha National Institute, Jatni, India
\item \Idef{org85}National Nuclear Research Center, Baku, Azerbaijan
\item \Idef{org86}National Research Centre Kurchatov Institute, Moscow, Russia
\item \Idef{org87}Niels Bohr Institute, University of Copenhagen, Copenhagen, Denmark
\item \Idef{org88}Nikhef, National institute for subatomic physics, Amsterdam, Netherlands
\item \Idef{org89}NRC Kurchatov Institute IHEP, Protvino, Russia
\item \Idef{org90}NRC Kurchatov Institute  - ITEP, Moscow, Russia
\item \Idef{org91}NRNU Moscow Engineering Physics Institute, Moscow, Russia
\item \Idef{org92}Nuclear Physics Group, STFC Daresbury Laboratory, Daresbury, United Kingdom
\item \Idef{org93}Nuclear Physics Institute of the Czech Academy of Sciences, \v{R}e\v{z} u Prahy, Czech Republic
\item \Idef{org94}Oak Ridge National Laboratory, Oak Ridge, Tennessee, United States
\item \Idef{org95}Ohio State University, Columbus, Ohio, United States
\item \Idef{org96}Petersburg Nuclear Physics Institute, Gatchina, Russia
\item \Idef{org97}Physics department, Faculty of science, University of Zagreb, Zagreb, Croatia
\item \Idef{org98}Physics Department, Panjab University, Chandigarh, India
\item \Idef{org99}Physics Department, University of Jammu, Jammu, India
\item \Idef{org100}Physics Department, University of Rajasthan, Jaipur, India
\item \Idef{org101}Physikalisches Institut, Eberhard-Karls-Universit\"{a}t T\"{u}bingen, T\"{u}bingen, Germany
\item \Idef{org102}Physikalisches Institut, Ruprecht-Karls-Universit\"{a}t Heidelberg, Heidelberg, Germany
\item \Idef{org103}Physik Department, Technische Universit\"{a}t M\"{u}nchen, Munich, Germany
\item \Idef{org104}Politecnico di Bari, Bari, Italy
\item \Idef{org105}Research Division and ExtreMe Matter Institute EMMI, GSI Helmholtzzentrum f\"ur Schwerionenforschung GmbH, Darmstadt, Germany
\item \Idef{org106}Rudjer Bo\v{s}kovi\'{c} Institute, Zagreb, Croatia
\item \Idef{org107}Russian Federal Nuclear Center (VNIIEF), Sarov, Russia
\item \Idef{org108}Saha Institute of Nuclear Physics, Homi Bhabha National Institute, Kolkata, India
\item \Idef{org109}School of Physics and Astronomy, University of Birmingham, Birmingham, United Kingdom
\item \Idef{org110}Secci\'{o}n F\'{\i}sica, Departamento de Ciencias, Pontificia Universidad Cat\'{o}lica del Per\'{u}, Lima, Peru
\item \Idef{org111}Shanghai Institute of Applied Physics, Shanghai, China
\item \Idef{org112}St. Petersburg State University, St. Petersburg, Russia
\item \Idef{org113}Stefan Meyer Institut f\"{u}r Subatomare Physik (SMI), Vienna, Austria
\item \Idef{org114}SUBATECH, IMT Atlantique, Universit\'{e} de Nantes, CNRS-IN2P3, Nantes, France
\item \Idef{org115}Suranaree University of Technology, Nakhon Ratchasima, Thailand
\item \Idef{org116}Technical University of Ko\v{s}ice, Ko\v{s}ice, Slovakia
\item \Idef{org117}Technische Universit\"{a}t M\"{u}nchen, Excellence Cluster 'Universe', Munich, Germany
\item \Idef{org118}The Henryk Niewodniczanski Institute of Nuclear Physics, Polish Academy of Sciences, Cracow, Poland
\item \Idef{org119}The University of Texas at Austin, Austin, Texas, United States
\item \Idef{org120}Universidad Aut\'{o}noma de Sinaloa, Culiac\'{a}n, Mexico
\item \Idef{org121}Universidade de S\~{a}o Paulo (USP), S\~{a}o Paulo, Brazil
\item \Idef{org122}Universidade Estadual de Campinas (UNICAMP), Campinas, Brazil
\item \Idef{org123}Universidade Federal do ABC, Santo Andre, Brazil
\item \Idef{org124}University of Cape Town, Cape Town, South Africa
\item \Idef{org125}University of Houston, Houston, Texas, United States
\item \Idef{org126}University of Jyv\"{a}skyl\"{a}, Jyv\"{a}skyl\"{a}, Finland
\item \Idef{org127}University of Liverpool, Liverpool, United Kingdom
\item \Idef{org128}University of Science and Techonology of China, Hefei, China
\item \Idef{org129}University of South-Eastern Norway, Tonsberg, Norway
\item \Idef{org130}University of Tennessee, Knoxville, Tennessee, United States
\item \Idef{org131}University of the Witwatersrand, Johannesburg, South Africa
\item \Idef{org132}University of Tokyo, Tokyo, Japan
\item \Idef{org133}University of Tsukuba, Tsukuba, Japan
\item \Idef{org134}Universit\'{e} Clermont Auvergne, CNRS/IN2P3, LPC, Clermont-Ferrand, France
\item \Idef{org135}Universit\'{e} de Lyon, Universit\'{e} Lyon 1, CNRS/IN2P3, IPN-Lyon, Villeurbanne, Lyon, France
\item \Idef{org136}Universit\'{e} de Strasbourg, CNRS, IPHC UMR 7178, F-67000 Strasbourg, France, Strasbourg, France
\item \Idef{org137}Universit\'{e} Paris-Saclay Centre d'Etudes de Saclay (CEA), IRFU, D\'{e}partment de Physique Nucl\'{e}aire (DPhN), Saclay, France
\item \Idef{org138}Universit\`{a} degli Studi di Foggia, Foggia, Italy
\item \Idef{org139}Universit\`{a} degli Studi di Pavia, Pavia, Italy
\item \Idef{org140}Universit\`{a} di Brescia, Brescia, Italy
\item \Idef{org141}Variable Energy Cyclotron Centre, Homi Bhabha National Institute, Kolkata, India
\item \Idef{org142}Warsaw University of Technology, Warsaw, Poland
\item \Idef{org143}Wayne State University, Detroit, Michigan, United States
\item \Idef{org144}Westf\"{a}lische Wilhelms-Universit\"{a}t M\"{u}nster, Institut f\"{u}r Kernphysik, M\"{u}nster, Germany
\item \Idef{org145}Wigner Research Centre for Physics, Hungarian Academy of Sciences, Budapest, Hungary
\item \Idef{org146}Yale University, New Haven, Connecticut, United States
\item \Idef{org147}Yonsei University, Seoul, Republic of Korea
\end{Authlist}
\endgroup
\end{document}